\documentclass[letterpaper]{aa}
\usepackage{graphicx}
\usepackage{graphics}
\usepackage{amssymb}

\begin{document}

   \title{High-Resolution Near-Infrared Speckle Interferometry 
and Radiative Transfer Modeling of the OH/IR star OH\,104.9+2.4${}^\star$}

\author{D.\,Riechers\inst1, 
Y.\,Balega\inst2,
T.\,Driebe\inst1,
K.-H.\,Hofmann\inst1, 
A. B. Men'shchikov\inst1{${}^,$}\inst3,
G.\,Weigelt\inst1}

   \institute{Max-Planck-Institut f\"ur Radioastronomie, Auf dem
              H\"ugel 69, D-53121 Bonn, Germany \\
              \email{riechers@mpifr-bonn.mpg.de, driebe@mpifr-bonn.mpg.de,
              weigelt@mpifr-bonn.mpg.de}
              \and
              Special Astrophysical Observatory, Nizhnij Arkhyz,
              Zelenchuk district, 
Karachai-Cherkessian Republic, Russia
              \and
              Institute for Computational Astrophysics, Saint Mary's University, Halifax, Canada
              }

\offprints{ D.~Riechers \\ \email{riechers@mpifr-bonn.mpg.de} \\
${}^\star$Based on data collected at the 6\,m BTA telescope 
of the Special Astrophysical Observatory in Russia}

  \date{Received 27 February 2004 ; accepted 18 May 2004}

\authorrunning{D.\,Riechers et al.}
\titlerunning{Near-Infrared Speckle Interferometry and Radiative Transfer Modeling of OH\,104.9+2.4}

\abstract{We present near-infrared speckle interferometry of the OH/IR star \object{OH\,104.9+2.4} 
in the $K'$ band obtained with the 6\,m telescope of the Special Astrophysical Observatory (SAO). 
At a wavelength of $\lambda = 2.12\,\mu$m the diffraction-limited resolution of 74\,mas was attained. 
The reconstructed visibility reveals a spherically symmetric, circumstellar dust shell (CDS) surrounding 
the central star. The visibility function shows that the stellar contribution to the total flux at 
$\lambda = 2.12\,\mu$m is less than $\sim\,$50\%, indicating a rather large optical depth of the CDS. 
The azimuthally averaged 1-dimensional Gaussian visibility fit yields a diameter of 47 $\pm$ 3\,mas (FHWM), 
which corresponds to 112 $\pm$ 13\,AU for an adopted distance of $D = 2.38 \pm 0.24$\,kpc. To determine the 
structure and the properties of the CDS of \object{OH\,104.9+2.4}, radiative transfer calculations using 
the code DUSTY were performed to simultaneously model its visibility and the spectral energy distribution (SED).
We found that both the ISO spectrum and the visibility of \object{OH\,104.9+2.4} can be well reproduced by 
a radiative transfer model with an effective temperature $T_{\rm eff} = 2500 \pm 500$\,K of the central 
source, a dust temperature $T_{\rm in} = 1000 \pm 200$\,K at the inner shell boundary $R_{\rm in} 
\simeq 9.1\,R_{\star} = 25.4\,$AU, an optical depth $\tau_{2.2 \mu \rm m} = 6.5 \pm 0.3$, and dust 
grain radii ranging from $a_{\rm min} = 0.005 \pm 0.003\,\mu$m to $a_{\rm max} = 0.2 \pm 0.02\,\mu$m with a 
power law $n(a) \propto a^{-3.5}$. It was found that even minor changes in  
$a_{\rm max}$ have a major impact on both the slope and the curvature of the visibility function, while 
the SED shows only minor changes. Our detailed analysis demonstrates the potential of dust shell modeling 
constrained by both the SED and visibilities. 

}

   \maketitle

   \section{Introduction}\label{intro}
\object{OH\,104.9+2.4} (\object{IRAS\,22177+5936}, \object{AFGL\,2885}, \object{NSV\, 25875}) 
is an OH/IR type II-A class star. These objects exhibit the maximum of their SED in the infrared 
(IR) around 6-10\,$\mu$m, while the 9.7\,$\mu$m silicate feature is found to be in absorption. 
In addition, they show strong OH maser line emission at 1612\,MHz/18\,cm (Habing \cite{hab96}) 
and weaker emission in SiO (43.1\,GHz and 86.2\,GHz), in $\rm H_2O$ (22.2\,GHz), and in the 
other hyperfine OH lines of the same transition (1665/1667\,MHz, 1720\,MHz) (Herman \& Habing 
\cite{her85}). OH/IR stars are long-period variables (pulsation periods between 500 and 3000\,d) 
of variability type Me, similar to long-period Mira stars. While the majority of OH/IR stars 
show a bolometric amplitude of typically $\sim\,1$\,mag, a very small fraction varies irregularly 
with a low amplitude or does not show any detectable variability. OH/IR stars are mostly low and 
intermediate mass (progenitor masses $M \leq 9\,M_{\rm \odot}$), oxygen-rich single stars evolving 
along the upper part of the asymptotic giant branch (AGB).  OH/IR stars extend the sequence of 
optical Mira variables towards longer periods, larger optical depths and higher mass-loss rates 
(Engels et al.\ \cite{eng83}, Lepine et al.\ \cite{lep95}). As a consequence of their high mass 
loss, OH/IR stars are surrounded by massive, optically and geometrically thick circumstellar 
envelopes composed of gas and dust. The strong maser emission reveals non-rotating CDS (Herman \& 
Habing \cite{her85}), which, in some cases, totally obscures the underlying star.

\indent Observations as well as theoretical models of the structure, dynamics, and evolution of 
the atmosphere and CDS of AGB stars have led to a detailed picture of these objects. According 
to Habing (\cite{hab96}), OH/IR stars are characterized by the following parameters: a variation 
of the bolometric flux $\Delta m_{\rm bol} > 0.7$\,mag, luminosities $L > 3 \cdot 10^3\,L_{\odot}$, 
effective temperatures $T_{\rm eff} < 3000$\,K, temperatures at the inner boundary of the CDS 
$T_{\rm in} \simeq 1000$\,K, optical depths $\tau_{\rm 9.7 \mu m} \simeq 10^{-2} - 10^1$, 
the silicate feature at this wavelength always in absorption, and high mass-loss rates 
($\dot{M} \simeq 10^{-7} - 10^{-4}\,M_{\odot}$/yr) at moderate outflow velocities ($v_{\rm e} 
\simeq 10$\,km/s). The AGB phase and therefore the phase of high mass loss typically lasts for 
some $10^5$\,yr. Lorenz-Martins \& de Araujo (\cite{lm97}) find  $T_{\rm eff} \simeq 1800 - 2400$\,K, 
$T_{\rm in} \simeq 650 - 1200$\,K, $\tau_{9.7 \mu \rm m} \simeq 7 - 17$, and the silicate feature 
typically in absorption for many OH/IR stars.

\indent The mass loss is driven by both large-amplitude pulsations and acceleration by radiation 
pressure. Therefore, correlations are found between the period and the terminal outflow velocity 
(e.g.\ Heske \cite{hes90}). 
Detailed hydrodynamical models show that, due to the passage of shocks 
generated by the stellar pulsation, the atmosphere is highly extended, thus enabling dust formation 
and the subsequent acceleration of matter (see, e.g., H\"ofner et al.\ \cite{hoefner03}, 
Schirrmacher et al.\ \cite{vasko03}, Jeong et al.\ \cite{jeong04}, and references therein).

\indent The assumption of a spherically symmetric dust shell is frequently motivated by the circularity 
of OH maser maps. High-spatial-resolution observations can give direct information about the extension 
and geometry of the CDS and therefore lead to more reliable constraints for the modeling of the 
circumstellar environment, supplementing the information from the SED. Measurements of the near-IR 
(NIR) visibility may be used to determine the radius of the onset of dust formation as well as to 
constrain the grain sizes (Groenewegen \cite{gw97}).

\indent \object{OH\,104.9+2.4} is a representative example of a type II-A OH/IR variable. From the 
variations of the 1612\,MHz OH maser, a period of $P = 1216 \pm 54$\,d was determined (Herman \& 
Habing \cite{her85}), although values up to 1620\,d have been found (Jones et al.\ \cite{jon90}). 
The radial outflow velocity of $v_{\rm e} = 15$\,km/s measured (te Lintel Hekkert et al.\ \cite{TLH91}) 
is in accordance with SiO and $\rm H_2O$ maser observations (Engels et al.\ \cite{eng86}). From maser 
phase lag measurements, an outer radius of the dust shell of $R_{\rm S} = (4.37 \pm 0.42) \cdot 
10^{16}$\,cm was determined, while the absolute distance was found to be $D = 2.38 \pm 0.24$\,kpc 
(Herman \& Habing \cite{her85}). A mass-loss rate of $\dot{M} = 5.58 \cdot 10^{-5}\,M_{\odot}/$yr 
was derived by Heske et al.\ (\cite{hfo90}) .

\indent The paper is organized as follows: in Sect.\ 2, results of NIR speckle interferometric 
observations of \object{OH\,104.9+2.4} are presented. Section 3 deals with the SED reconstruction. 
The approach for the radiative transfer modeling (RTM), as well as the strategy of simultaneous 
modeling of SED and near-infrared visibility is outlined in Sect.\,4.1. The method described was 
recently applied to other evolved stars, such as \object{Red Rectangle} (Men'shchikov et al.\ 
\cite{men98}), \object{AFGL\,2290} (Gauger et al.\ \cite{gau99}), \object{IRC\,+10420} (Bl\"ocker 
et al.\ \cite{blo99}), \object{NML Cyg} (Bl\"ocker et al.\ \cite{blo01}) and \object{CIT\,3} 
(Hofmann et al.\ \cite{Hof01}). Sections~4.2-4.4 give details of the modeling results, and Sect.\ 5 
closes the discussion with a summary of the results, a conclusion and an outlook. 
Additional results will be presented in a subsequent publication (Riechers et al.\ \cite{rie04}).

   \section{Speckle interferometry observations and data reduction}\label{obsres}
\begin{figure*}
      \begin{center}
\begin{minipage}[t]{65mm}{
\resizebox{60mm}{!}{
\includegraphics[angle=0]{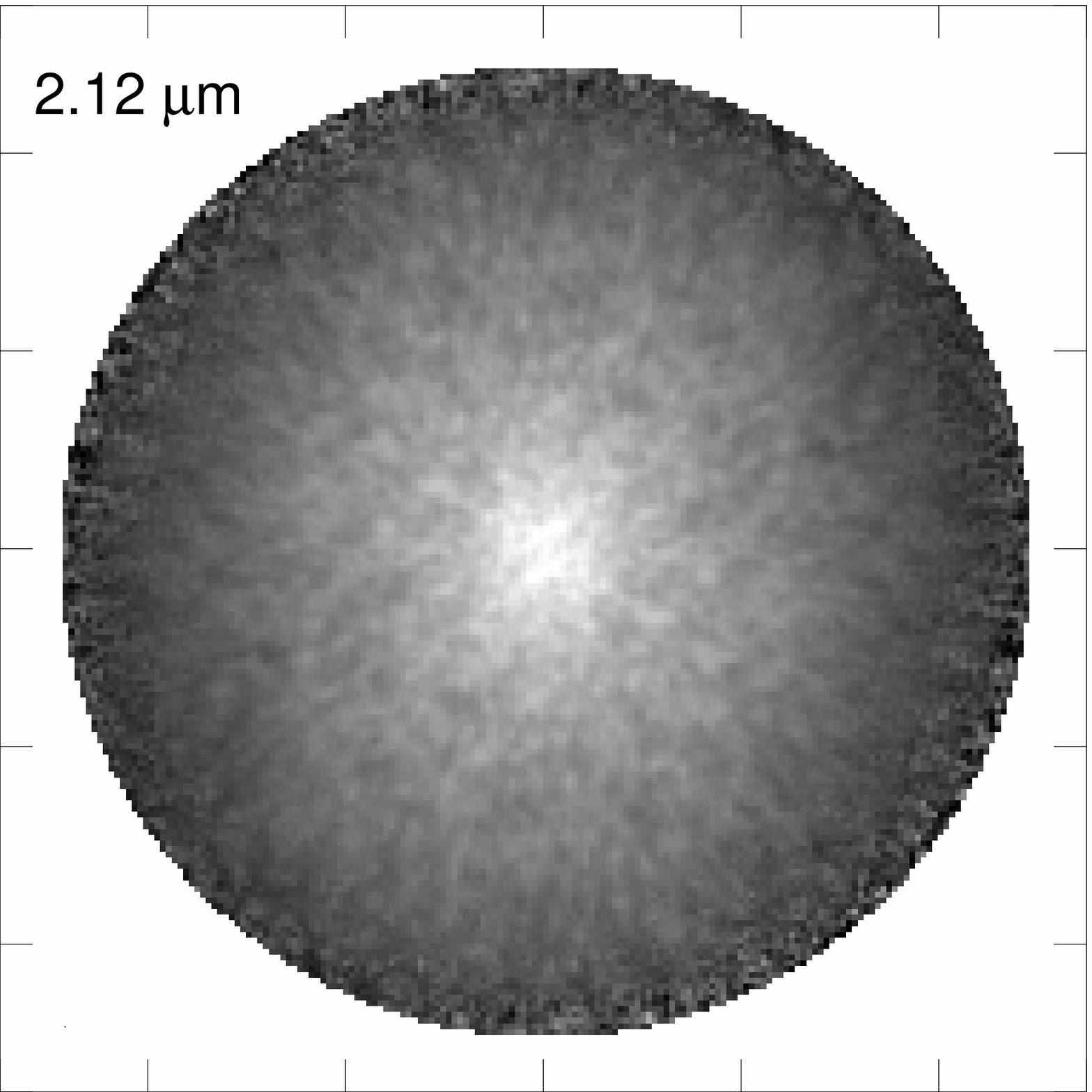}}
}
\end{minipage}
\begin{minipage}[t]{60mm}{
\resizebox{60mm}{!}{
\includegraphics[angle=0]{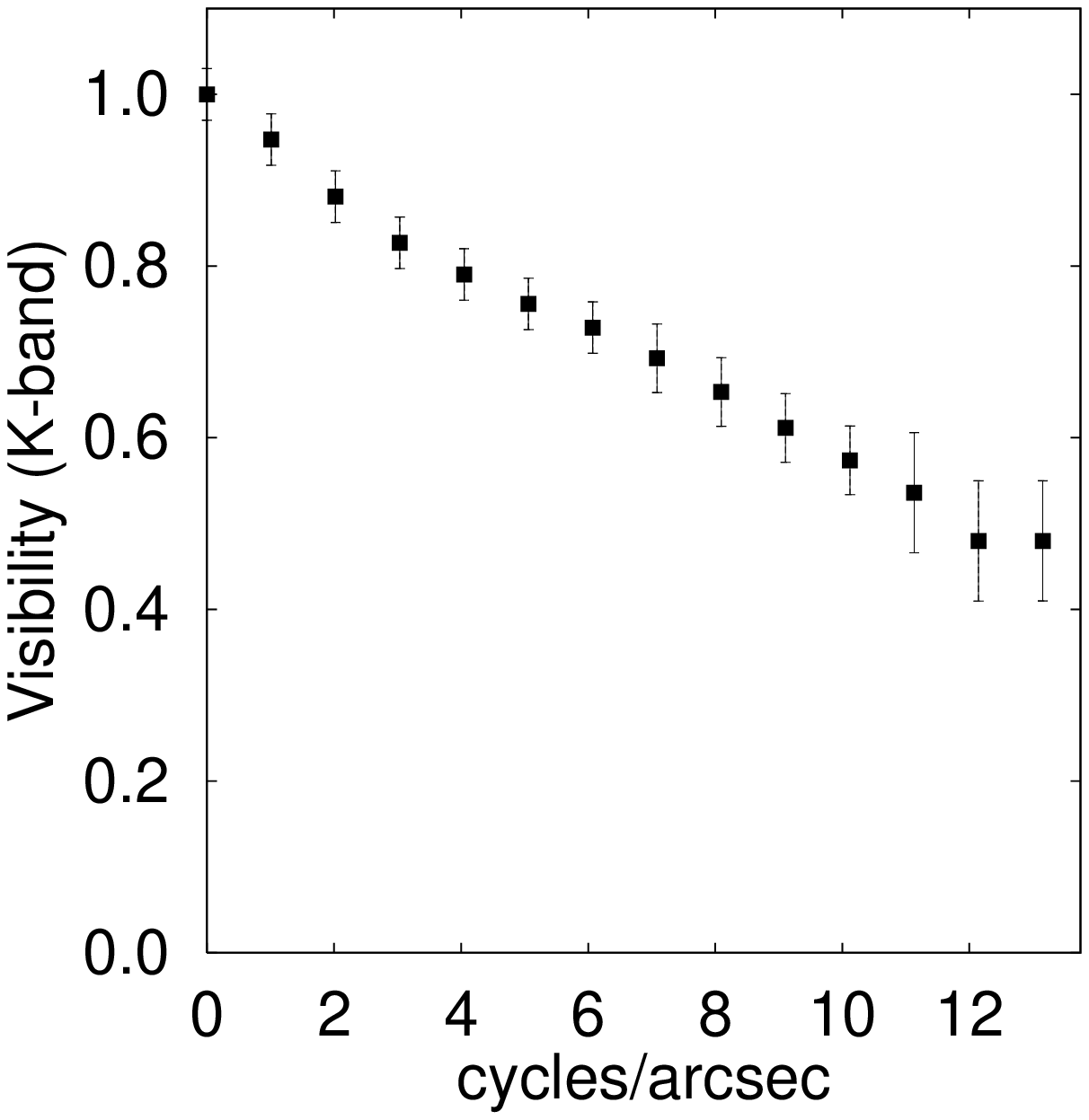}}
}
\end{minipage}
\begin{minipage}[t]{40mm}{
\vspace*{-50mm}

\caption{
Observations of \object{OH\,104.9+2.4} on September 22, 2002, with the SAO 6\,m telescope. 
\textbf{Left}: 2-dimensional 2.12\,$\mu$m visibility function of \object{OH\,104.9+2.4}. 
\textbf{Right}: Azimuthally averaged visibility of \object{OH\,104.9+2.4} in
the $K^{\prime}$ band. 
}
}
\end{minipage}\hfill
      \end{center}
   \label{f1}
\end{figure*}

The $K'$-band speckle interferograms of \object{OH\,104.9+2.4} were obtained with the 
Russian 6\,m telescope of the Special Astrophysical Observatory (SAO) on September 22, 2002. 
The data were recorded with our HAWAII speckle camera through interference filters
with a center wavelength of 2.12\,$\mu$m and a bandwidth of 0.21\,$\mu$m ($K^{\prime}$ band).
Additional speckle interferograms were taken for three unresolved reference stars 
(\object{GJ\,105.5}, \object{G\,77-31}, and \object{HD\,22686}). With a pixel size of 27\,mas 
and a seeing of 1.9 arcsec ($1.5\,\cdot\,$FWHM of centered long exposure), 444+727\,object 
frames and 690+792\,frames of the reference stars were taken, each with an exposure time of 268\,ms. 
These interferograms were used to compensate for the speckle interferometry transfer function. 
The visibility function of \object{OH\,104.9+2.4} was derived from the speckle interferograms using 
the speckle interferometry method (Labeyrie \cite{l70}). The reconstructed two-dimensional visibility 
at 2.12\,$\mu$m and the azimuthally averaged visibility profile is shown in Fig.\,\ref{f1}. 
The azimuthally averaged visibility decreases steadily to values below $\sim$\,0.50 at the 
diffraction cut-off frequency (13.5\,cycles/arcsec). Thus, the CDS is almost fully resolved, 
and the contribution of the unresolved component to the monochromatic flux at  $\lambda = 2.12\,\mu$m 
is less than $\sim$\,50\,\%, indicating a rather large optical depth at this wavelength. In order 
to give a first, rough estimate for the apparent diameter of the dust shell, the azimuthally 
averaged visibility was fitted with a Gaussian and a uniform disk (UD) center-to-limb variation. 
We obtain a Gaussian fit FWHM diameter of 47 $\pm$ 3\,mas, which corresponds to a diameter 
of 112 $\pm$ 13\,AU for an adopted distance of $D = 2.38 \pm 0.24$\,kpc (Herman \& Habing \cite{her85}). 
The UD diameter is found to be 73 $\pm$ 5\,mas, corresponding to 173 $\pm$ 22\,AU. The quality 
of the visibility fits with both the Gaussian model and the UD are rather poor, as the shape 
of the observed visibility strongly deviates from that of a Gaussian or UD visibility. 
From our observations, no major
deviation from spherical symmetry was detected, in accordance with the analysis of 
OH maser maps (Welty et al.\ \cite{wel87}). 
Given the accuracy and resolution of our visibility measurement, 
we find that the diameter ratio between major and minor axis is less than 8\,\%.
Therefore, we restricted the further discussion and the modeling itself
to the azimuthally averaged 1-dimensional visibility and the usage
of a 1D-radiative transfer code.

   \section{Spectral energy distribution}\label{sed}
\begin{figure*}
      \begin{center}
\begin{minipage}[t]{115mm}{
\resizebox{115mm}{!}{
\includegraphics[angle=0]{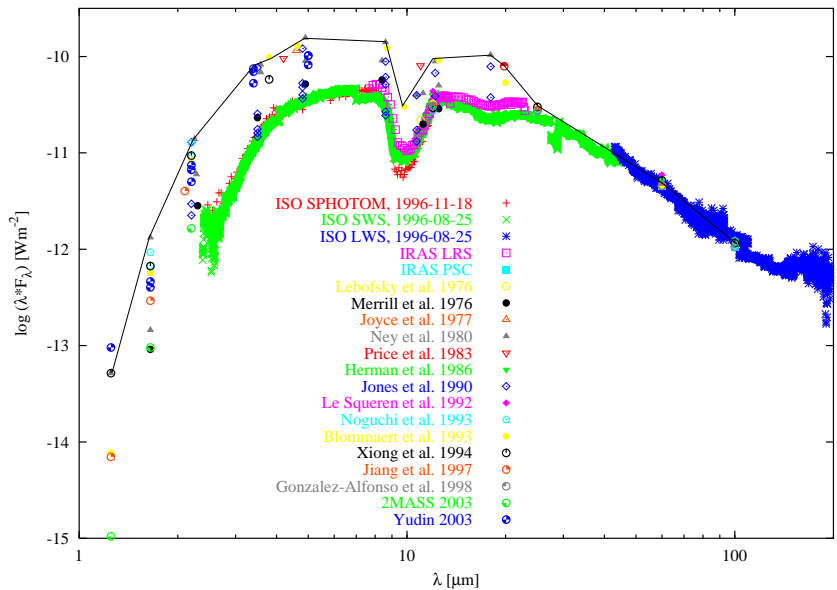}
}
}
\end{minipage}
\begin{minipage}[t]{60mm}{
\vspace*{-80mm}

\caption{
SED of \object{OH\,104.9+2.4}. The ISO SED appears to be taken at near-minimum pulsation phase 
($\Phi_{\rm ISO} = 0.5$). The black line depicts a maximum envelope for the SED, which was used 
to derive the bolometric flux at maximum phase. 
}
}
\end{minipage}\hfill
      \end{center}
\vspace*{-8mm}

   \label{f2}
\end{figure*}

\begin{figure*}
      \begin{center}
\begin{minipage}[t]{100mm}{
\resizebox{100mm}{!}{
\includegraphics[angle=-90]{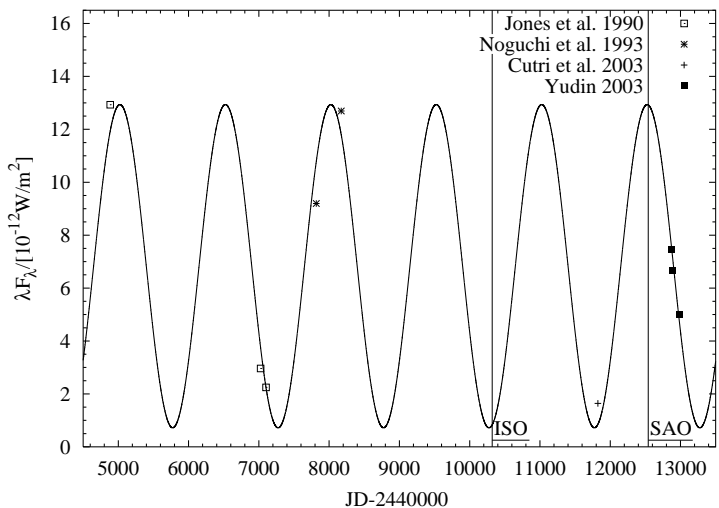}}
}
\end{minipage}\hfill
\begin{minipage}[t]{75mm}{
\caption{
Recent observations of \object{OH\,104.9+2.4} at $\lambda = 2.2\,\mu$m. Measurements by Jones 
et al.\ (\cite{jon90}), Noguchi et al.\ (\cite{nog93}), Cutri et al.\ (\cite{2MASS}) and Yudin 
(\cite{yud03}) are shown. A 4-parameter cosine function is fitted (least-squares method) to 
the data and provides a pulsation period of $P = 1500 \pm 11\,$d. The left vertical line indicates 
the date of the ISO measurements and substantiates the assumption of $\Phi_{\rm ISO} = 0.5$. The 
right vertical line indicates the date of the SAO measurenents and allows us to derive 
$\Phi_{\rm SAO} \simeq 0$.
}
}
\end{minipage}
      \end{center}
\vspace*{-3mm}

   \label{f2x}
\end{figure*}

Photometric and spectroscopic data in the IR for \object{OH\,104.9+2.4} have been reported, for example, 
by Lebofsky et al.\ (\cite{leb76}), Merrill et al.\ (\cite{mer76}), Joyce et al.\ (\cite{joy77}), 
Ney et al.\ (\cite{ney80}), Price et al.\ (\cite{pri83}), Knacke et al.\ (\cite{kna85}), 
Herman et al.\ (\cite{her86}), Cobb et al.\ (\cite{cob87}), Jones et al.\ (\cite{jon90}), 
Le Squeren et al.\ (\cite{squ92}), Noguchi et al.\ (\cite{nog93}), Blommaert et al.\ (\cite{blo93}), 
Xiong et al.\ (\cite{xio94}), Jiang et al.\ (\cite{jia97}), Gonzalez-Alfonso et al.\ (\cite{gon98}), 
and Cutri et al. (\cite{2MASS}). \object{OH\,104.9+2.4} was observed by IRAS 
(Infrared Astronomical Satellite) in 1983 and by ISO (Infrared Space Observatory) in 1996. 
In Fig.\,\ref{f2}, the SED of \object{OH\,104.9+2.4} is shown. In addition to the cited data, 
recent data obtained by Yudin (\cite{yud03}) at the Crimean Astrophysical Observatory (CAO) on 
August 14, 2003, September 08, 2003, and December 10, 2003, in the $J$, $H$, $K$, $L$, and 
$M$ bands (Tab.\,\ref{tab-1}) are shown. The reference star for these observations was 
\object{BS\,8465} ($J = 0.97^{\rm m}$, $H = 0.28^{\rm m}$, $K = 0.11^{\rm m}$, $L = 0.00^{\rm m}$, 
$M = 0.20^{\rm m}$). \object{OH\,104.9+2.4} shows strong long-term SED variations (see Fig.\,\ref{f2}). 
The ISO data were most likely taken at minimum pulsation phase, as all other data points lie 
above the ISO spectrum. Therefore, we adopt $\Phi_{\rm ISO} \simeq 0.5$. From the ISO data, 
a near-minimum bolometric flux of $F_{\rm bol}^{\rm ISO} = 6.99 \cdot 10^{-11}\,\rm{W/m^2}$ can be 
derived. In order to find the bolometric flux at the phase of the SAO observations, a bolometric 
flux at maximum phase has to be determined first. To obtain this bolometric flux, we defined a 
maximum envelope representing the maximum-luminosity SED as depicted by the black line in Fig.\,\ref{f2}, 
which takes into account the wavelength dependence of the pulsation amplitude, and obtain 
$F_{\rm bol}^{\rm max} = 2.30 \cdot 10^{-10}\,\rm{W/m^2}$. With $K$-band data measured by 
Jones et al.\ (\cite{jon90}), Noguchi et al.\ (\cite{nog93}), Cutri et al.\ (\cite{2MASS}), 
and Yudin (\cite{yud03}) and a simple cosine-like pulsation model, we derived 
a pulsation period of $P = 1500 \pm 11\,$d, applying a 4-parameter least-squares fit 
($f(x) = a \cdot \cos(2\pi x/P-b)+c$), as shown in Fig.\,\ref{f2x}. 
We would like to point out, that this model is not physically accurate, 
but a suitable choice regarding the low number of data points.
With this model fit, we find $\Phi_{\rm SAO} \simeq 0$. Therefore, 
$F_{\rm bol}^{\rm SAO} = 2.30 \cdot 10^{-10}\,\rm{W/m^2}$ can be assumed, corresponding 
to a factor of $\sim$\,3.3 relative to the flux at $\Phi_{\rm ISO}$. This difference in 
$F_{\rm bol}$ has to be taken into account in the modeling process.

\indent \object{OH\,104.9+2.4} is highly reddened by the interstellar medium (ISM) and its CDS. 
The interstellar reddening was taken into account by adopting the method of Savage \& Mathis 
(\cite{SM79}) with $A_V = 3.1 \cdot E(B-V)$. For a distance of $D = 2.38 \pm 0.24\,$kpc, we found 
$A_V = 4.09$. This correction was applied to all models as described by Men'shchikov et al.\ 
(\cite{men02}). 

\begin{table*}
\caption{
Infrared photometry of \object{OH\,104.9+2.4}. The observations were carried out 
by Yudin (\cite{yud03}). All observations are given in magnitudes.}
 
\label{tab-1} 
\centering
\vspace*{-3mm}

\begin{tabular}{ c c r r r r r r }\hline
Date & Julian Date & $J$ & $H$ & $K$ & $L$ & $M$ 
\\ \hline
Aug 14, 2003 & 2452866.50 &  & 9.04 & 5.20 & 1.16 & -0.06 \\
Sep 08, 2003 & 2452891.42 & 11.53 & 9.19 & 5.32 & 1.24 & -0.06 \\
Dec 10, 2003 & 2452984.20 & & & 5.63 & 1.54 & 0.18 \\
\hline
\end{tabular}\\
\end{table*}

   \section{Dust shell models}\label{models}
\subsection{The radiative transfer code}

The radiative transfer calculations were performed using the publicly available code 
DUSTY (Ivezi{\' c} et al.\ \cite{IE99}). The main assumptions involved are (for 
single-shell models): $(i)$ constant spherical symmetry with $\rho(r) \propto r^{-n}$, $(ii)$ 
a constant dust-to-gas ratio, $(iii)$ a constant dust opacity and size distribution, 
$(iv)$ radiative equilibrium, $(v)$ the same temperature $T_{\rm d}(a,r) = T_{\rm d}(r)$
for all dust grains, and $(vi)$ isotropic scattering. 
With these assumptions, the only relevant property of the input radiation is its 
spectral shape $f_{\rm \lambda} = F_{\rm \lambda}/F$. The overall optical depth of the 
dust envelope at some reference wavelength $\tau_{\rm \lambda}$ fixes the solution, 
allowing only for normalized density distributions $\rho_{\rm d}$ describing the spatial 
variation of the dust, and the wavelength dependence of the 
optical properties of the dust grains. 

\indent The DUSTY formulation of the radiative transfer problem for a dusty envelope is 
very suitable for modeling of IR observations, since it minimizes the number of independent 
model input parameters. The only parameters that have to be specified are $(i)$ the spectral 
shape of the central source of radiation $f_{\rm \lambda}$ (i.e.\ the variation of the 
monochromatic flux with wavelength), $(ii)$  the absorption/emission and scattering 
efficiencies of the grains (i.e.\ the chemical composition and the grain-size distribution), 
$(iii)$ the normalized density distribution of the dust $\rho(r)/\rho_{\rm in}$, $(iv)$ the 
relative thickness of the dust shell (i.e.\ the ratio of outer to inner dust shell radius, 
$r_{\rm out}/r_{\rm in}$), $(v)$ the dust temperature $T_{\rm in}$ at the inner boundary 
$r_{\rm in}$, and $(vi)$ the overall optical depth at a reference wavelength $\tau_{\rm \lambda}$. 
For a given set of input parameters, DUSTY iteratively determines the radiation field and 
the dust temperature distribution (generalized output quantities) by solving the equation 
of radiative transfer. 

\subsection{Previous modeling of \object{OH\,104.9+2.4}}

There are two previous studies which deal with dust shell modeling of \object{OH\,104.9+2.4}. 
The first dust shell models were reported by Ivezi{\' c} et al.\ (\cite{IE95}) using a 
previous version of the DUSTY code. It was mainly based on IRAS data. The authors 
found $F_{\rm bol} = 7.6 \cdot 10^{-11}$\,W/$\rm m^2$, $L = 1.19 \cdot 10^4\,L_{\odot}$ 
for $D = 2.30$\,kpc, $v_{\rm e} = 15.1$\,km/s, $\tau_{2.2\,\mu{\rm m}} = 1.12$, and 
$\dot{M} = 3.87 \cdot 10^{-5}\,M_{\odot}$/yr. 

\indent A recent two-component dust shell model was reported by Suh (\cite{Suh02}), 
using only the ISO SWS/LWS spectra as observational constraints. The author introduced 
a combined envelope/disk model to improve his fit to the observational data. The 
reference optical depth for his best-fitting model was $\tau_{10\,\mu{\rm m}} = 10.0$. 
To date, there has not been any modeling of \object{OH\,104.9+2.4} that uses 
observational constraints from both the SED and visibility measurements.

\subsection{The best-fitting model}

For the analysis of OH\,104.9+2.4 based on the radiative transfer modeling, $\sim 10^6$ 
models (discussed in Sect.\,4.4) were calculated. To simultaneously fit the SED and 
visibility data (i.e.\ with identical model parameters), $F_{\rm bol}$ had to be 
adjusted for the epoch of the visibility measurements, since the ISO spectrum and our 
visibility data were taken at different pulsation phases of \object{OH\,104.9+2.4} 
(see Fig.\,\ref{f2x}). Therefore, the best-fitting model (Figs.\,\ref{f3} and \ref{f15}) 
consists of an SED model assuming the 1996 ISO flux and a visibility model assuming 
the 2002 SAO flux. The results are all summarized in Tab.\,\ref{tab-7}.

\subsubsection{SED fit}

\begin{figure*}
      \begin{center}
\resizebox{8.95cm}{!}{\includegraphics[angle=-90]{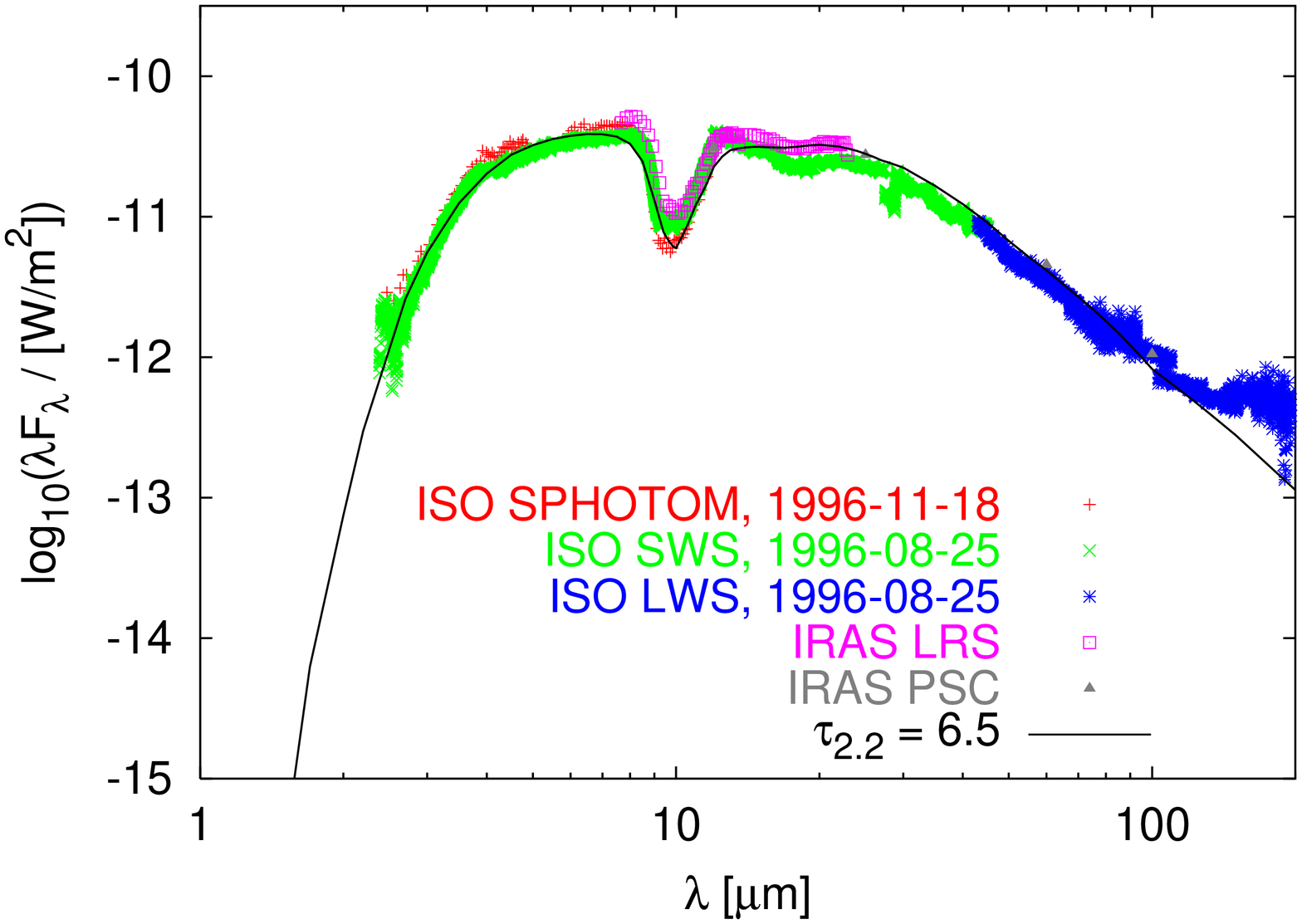}}
\resizebox{8.95cm}{!}{\includegraphics[angle=-90]{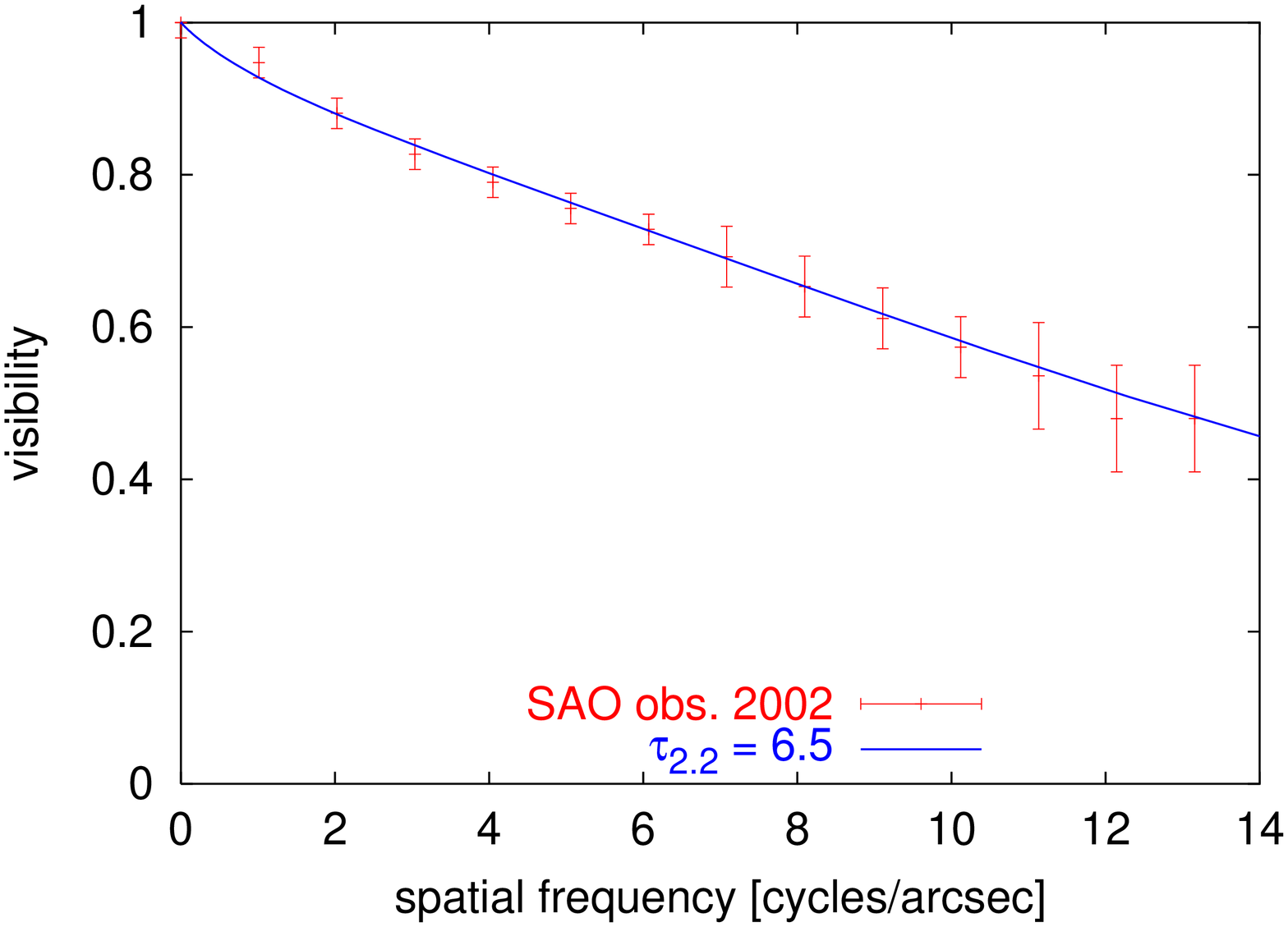}}
   \end{center}
\vspace*{-5mm}

\caption{
SED ({\bf left}) and visibility ({\bf right}) of \object{OH\,104.9+2.4} for our 
best-fitting model. As described in the text, the bolometric flux used for the SED 
is a factor of 3.3 lower than the one used for the SED, in order to take into account 
the different epochs of the ISO and SAO observations.
}
   \label{f3}
\end{figure*}

The final model that fits the ISO SED data of \object{OH\,104.9+2.4} best is presented 
in Fig.\,\ref{f3} (left panel). The black-body (BB) effective temperature of the central 
source of radiation is $T_{\rm eff} = 2500$ K. In addition, $T_{\rm in} = 1000$\,K, 
$\rho(r)/\rho_{\rm in} \propto r^{-2.0}$ and $r_{\rm out}/r_{\rm in} = 10^5$. 
Assuming the overall influence of the dust formation zone on the SED to be rather small, 
a standard $n(a) \propto a^{-3.5}$ grain-size distribution (GSD) function (Mathis, Rumpl, 
\& Nordsieck\ \cite{mrn77}, hereafter MRN) was chosen, with $a$ ranging between  
$a_{\rm min} = 0.005\,\mu$m and $a_{\rm max} = 0.2\,\mu$m. A dust composition with 95\,\% 
of the optical constants for warm silicates from Ossenkopf, Henning, \& Mathis (\cite{oss92}, 
hereafter OHM) and 5\,\% astronomical silicates from Draine \& Lee (\cite{DL84}, hereafter DL) 
was found to give the best results. The OHM warm, oxygen-deficient silicates are based on 
observational determinations of the opacities of circumstellar silicates as well as laboratory 
data and also take into account the effects of inclusions (amorphous and crystalline 
substructures as well as other materials) on the complex refractive index, especially at 
$\lambda = 8\,\mu$m. The reference optical depth in our best-fitting model is 
$\tau_{2.2\,\mu{\rm m}} = 6.5$, corresponding to $\tau_{9.7\,\mu{\rm m}} = 14.0$ for 
this model at the prominent silicate absorption feature. The observational value of 
$F_{\rm bol} = 6.99 \cdot 10^{-11}$\,W/$\rm m^2$ from 
the ISO data can be reproduced well by our model. 

\indent From the shortest wavelength at $\lambda = 2.38\,\mu$m up to $\lambda = 196\,\mu$m, 
the model provides a good fit to the observations. The location, shape, and strength of the 
silicate feature at $\lambda = 9.7\,\mu$m caused by SiO stretching vibrations is fairly well 
reproduced by the given composition. In the $\lambda \simeq 18\,\mu$m region, where 
$\rm SiO_2$ bending vibrations dominate the spectrum, there is a noticeable deviation from 
the observational data. Above $\lambda \simeq 100\,\mu$m, the model falls off too steeply to 
fit the data. On the one hand, this may be caused by deteriorating data quality towards the 
long wavelengths. On the other hand, the discrepancy at longer wavelengths may be due to the 
fact that DUSTY uses a GSD as input but internally averages dust properties over 
the entire size distribution.

\subsubsection{Visibility fit}

From Fig.\,\ref{f2} it is obvious that the ISO SED and the SAO visibility data were 
taken at different pulsation phases. While the ISO observations were made at near-minimum 
phase, the SAO data were taken at near-maximum phase. Therefore, it is quite natural that 
the two constraints cannot be fulfilled by one model alone with the same $F_{\rm bol}$. 
A factor of $\sim$\,3.3 in terms of the flux at ISO phase has to be introduced to the 
visibility model if it is to be comparable to the SED model, as outlined in Sect.\,3. 
If we combine the best-fitting SED model with $F_{\rm bol} = 6.99 \cdot 10^{-11}$\,W/$\rm m^2$ 
from the previous subsection with a corresponding visibility model with $F_{\rm bol} = 2.30 
\cdot 10^{-10}$\,W/$\rm m^2$, we obtain a very good simultaneous model for the SED and the 
visibility of \object{OH\,104.9+2.4}, as shown in Fig.\,\ref{f3}. 
However, it should be stated that the physical conclusions that can be made based on 
this simultaneity itself are limited because of the simplifications involved in the phase 
alignment and the simplicity of the pulsation model itself (only flux scaling, assumption
of time-independent $T_{\rm eff}$ and $R_{\star}$). Therefore, all futher derived quantities 
are given for the phase of the ISO measurements unless explicitly stated otherwise.

\begin{table*}

\caption{
The derived physical parameters of \object{OH\,104.9+2.4} as provided by the best-fitting model. 
}

\label{tab-7}
\begin{center} 
\begin{tabular}{ l l c l }\hline
Parameter & Value & Varied range & Impact on SED/visibility
\\ \hline
Effective temperature (black-body) & $T_{\rm eff} = 2500 \pm 500\,$K & $1500 \dots 5000\,$K & very weak/weak \\
Temperature at inner CDS boundary & $T_{\rm in} = 1000 \pm 200\,$K & $500 \dots 3000\,$K & moderate/moderate \\
Density profile within the CDS & $\rho (r) \propto r^{-n}$ with $n = 2.0 \pm 0.1$ & $-1 \dots 4^{\star}$ 
& moderate/weak \\
Relative CDS thickness & $\frac{r_{\rm out}}{r_{\rm in}} = 10^p$ with $p = 5^{+\infty}_{-2}$ & $1 \dots 7$ 
& ${}^{\rm very\,\,weak/very \,\,weak\,\,(p\ga3),}_{\rm moderate/weak\,\,(p\la3)}$ \\
Dust grain distribution function & MRN, $n(a) \propto a^{-q}$ with $q = 3.5$ & $2 \dots 5^{\star\star}$ 
& moderate/strong \\
Minimum grain size & $a_{\rm min} = 0.005 \pm 0.003\,\mu$m & $0.0005 \dots 0.5\,\mu$m & very weak/moderate \\
Maximum grain size & $a_{\rm max} = 0.2 \pm 0.02\,\mu$m & $0.05 \dots 500\,\mu$m & weak/very strong \\
Dust types & 95\,\% OHM warm silicates & & very strong/weak \\
& 5\,\% DL astronomical silicates & & \\
Optical depth & $\tau_{0.55\,\mu\rm{m}} = 158 \pm 7$ & & \\
& $\tau_{2.2\,\mu\rm{m}} = 6.5 \pm 0.3$ & $0.5 \dots 100$ & moderate/strong \\
& $\tau_{9.7\,\mu\rm{m}} = 14.0 \pm 0.6$ & & \\
Radius & $R_{\star} \simeq 600\,R_{\odot} = 2.79\,$AU & & \\
Radius of inner CDS boundary & $R_{\rm in} = 9.1\,R_{\star} = 25.4\,$AU & & \\
Radius of $I = 10^{-10} I_{\rm max}$ & $R_{\rm out}^{-10} = 5 \cdot 10^3\,R_{\star} = 1.4 \cdot 10^4\,$AU & & \\
Outer radius of the model & $R_{\rm out} = 9.1 \cdot 10^5\,R_{\star} = 2.5 \cdot 10^6\,$AU & & \\
Mass-loss rate & $\dot{M} = 2.18 \cdot 10^{-5}\,M_{\odot}/$yr & & \\
\hline
\end{tabular}
\end{center}
${}^{\star}$: parameter variation includes broken power laws, superwind models
and models with radiatively driven winds\\
${}^{\star\star}$: parameter variation includes test of alternative grain size 
distributions (see text)\\
\vspace*{-5mm}

\end{table*}

\subsubsection{Further model results}

\begin{figure*}
      \begin{center}
\resizebox{78mm}{!}{\includegraphics[angle=-90]{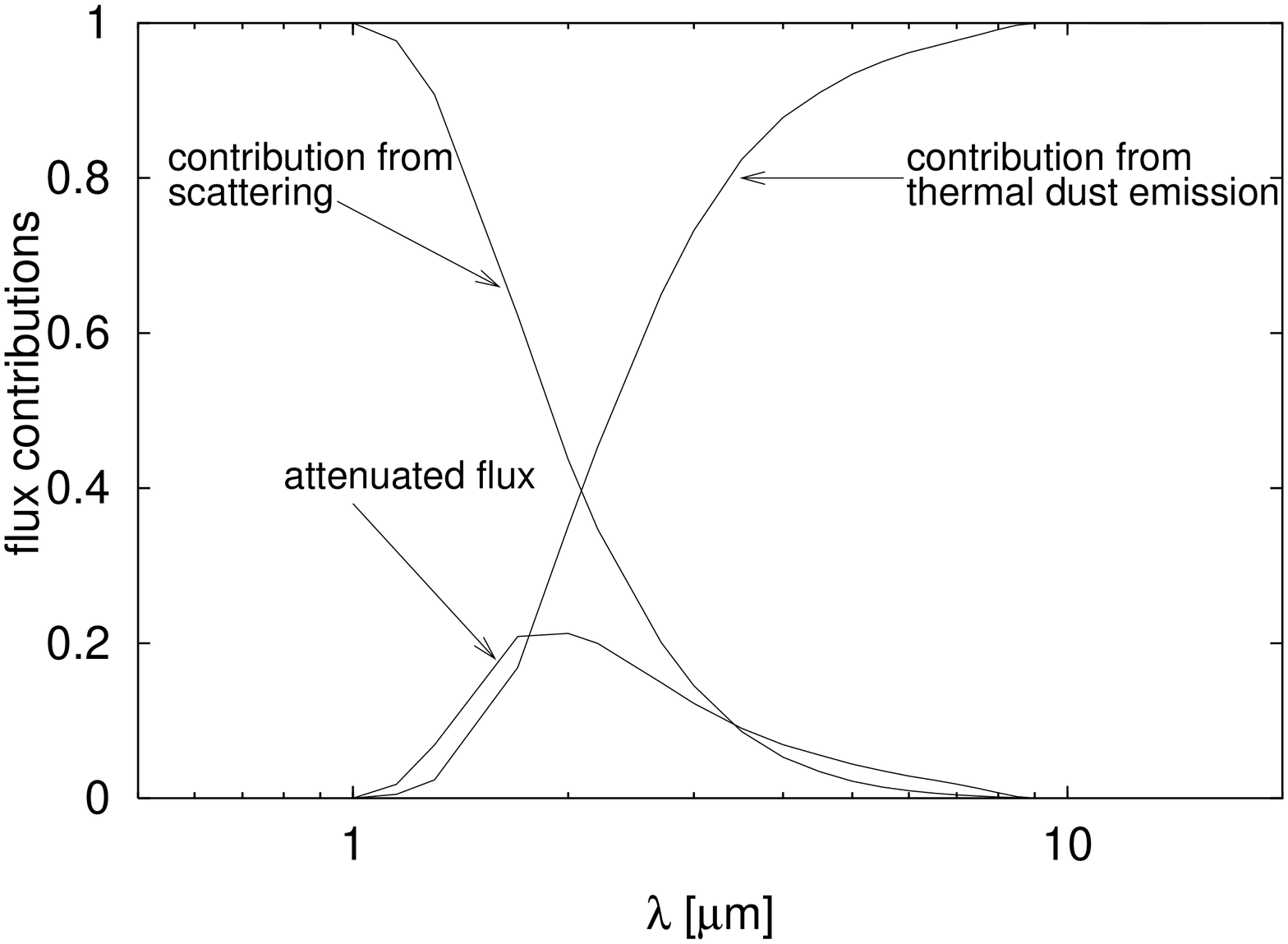}}
\resizebox{78mm}{!}{\includegraphics[angle=-90]{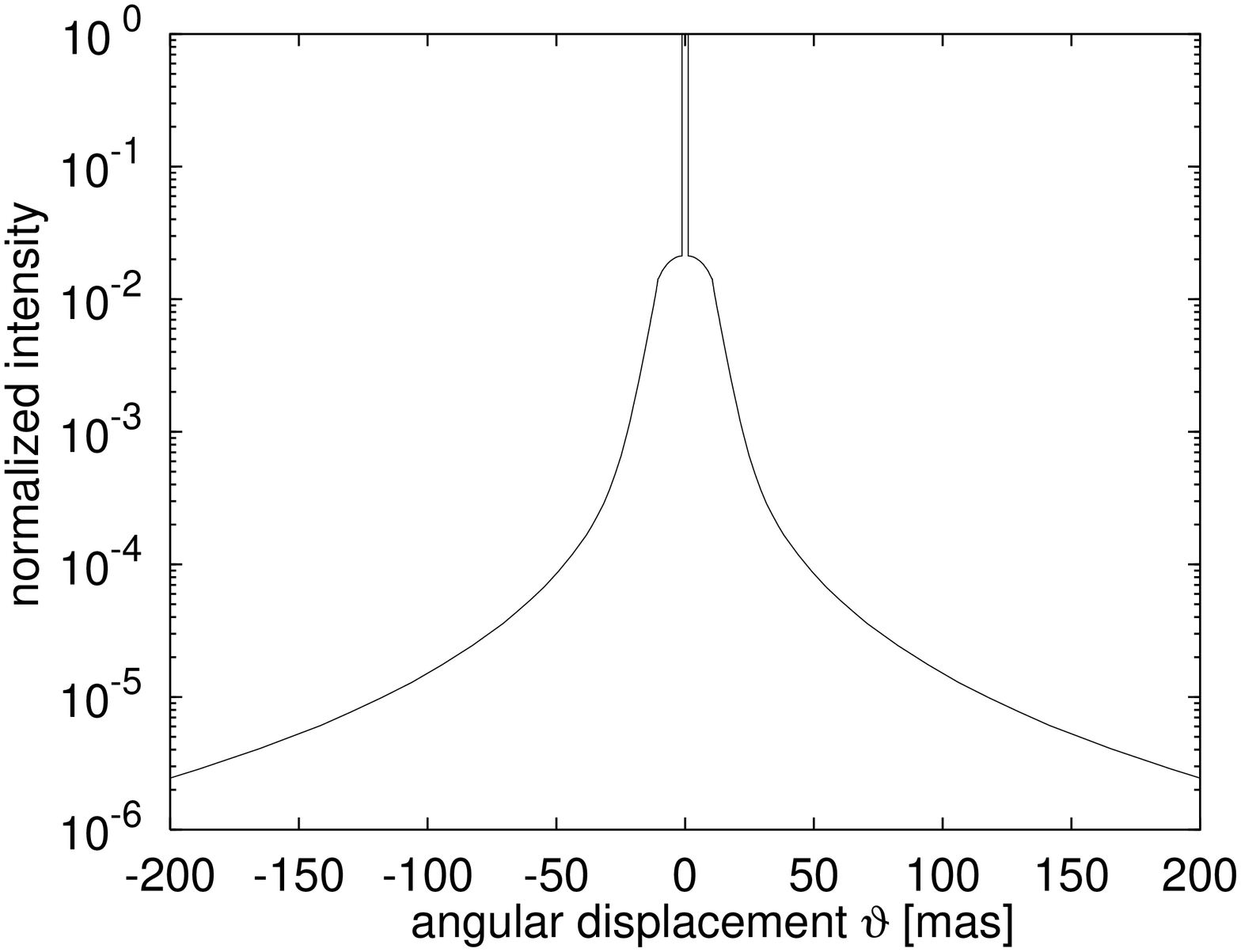}}
\resizebox{78mm}{!}{\includegraphics[angle=-90]{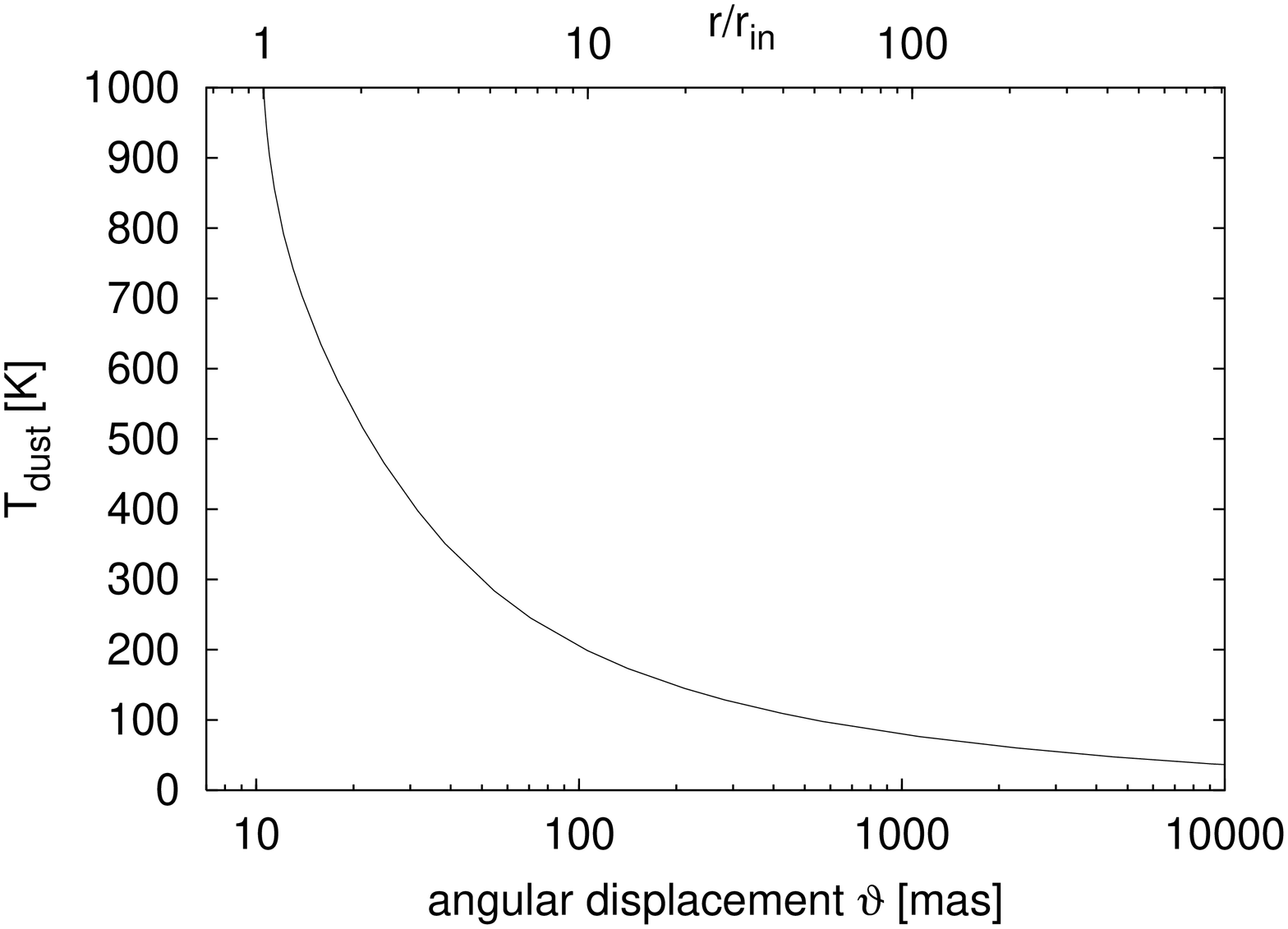}}
\resizebox{78mm}{!}{\includegraphics[angle=-90]{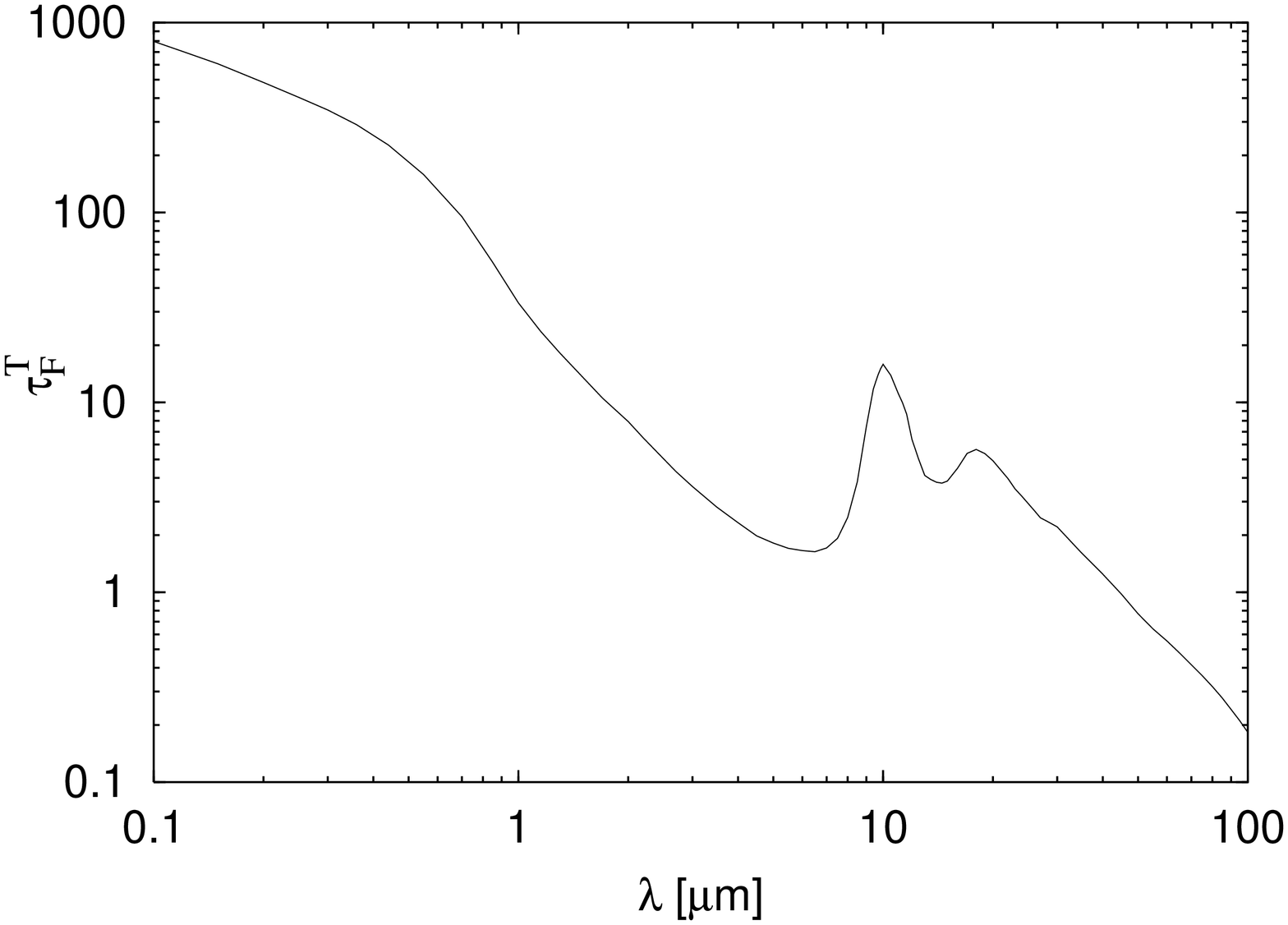}}
      \end{center}
\vspace*{-5mm}

\caption{
Modeling results of \object{OH\,104.9+2.4} for our best-fitting model (see Tab.\,\ref{tab-7}). 
\textbf{Top left}: Fractional contributions to the total flux. The contribution from 
direct stellar light is small in comparison to the contribution of the processes in the CDS.
\textbf{Top right}: The normalized intensity profile at 2.13\,$\mu$m. 
The sharp central peak corresponds to the central source of radiation. 
\textbf{Bottom left}: Dust temperature as a function of angular distance (lower axis) 
and of the radius in units of the inner dust shell radius (upper axis). The point where 
$T_{\rm dust} = 1000$\, K indicates the inner radius of the dust shell ($\frac{r}{r_{\rm in}} = 1$).
\textbf{Bottom right}: Wavelength depencence of the total optical depth.
}
   \label{f15}
\end{figure*}

The top left panel of Fig.\,\ref{f15} shows the fractional flux contributions of 
the emerging stellar radiation, the scattered radiation, and thermal dust emission 
as a function of wavelength, for our best-fitting model. At 2.2\,$\mu$m, the flux 
is dominated by thermal dust emission (45\,\%) and scattering (35\,\%). Direct 
stellar light (attenuated flux) contributes only 20\,\% at this wavelength. Thus, 
scattering contributes nearly to the same extent to the radiation of the CDS in 
the $K$ band as thermal dust emission. At shorter wavelengths scattering dominates, 
while dust emission is the major flux contribution at wavelengths longer than 3\,$\mu$m. 
The attenuated flux from the central source plays a minor role at all wavelengths under 
consideration (0.01\,$\mu$m - 0.03\,m) due to nearly total obscuration.

\indent The top right panel of Fig.\,\ref{f15} shows the normalized intensity 
distribution at 2.13 $\mu$m as a function of angular distance. The barely resolved 
central peak corresponds to the central star. At the given optical depth of 
$\tau_{2.2\,\mu{\rm m}} = 6.5$, there is no evidence for limb-brightening effects 
at the inner boundary, due to the extreme optical thickness of the CDS. 
If we define an outer radius of the dust shell $R_{\rm out}^{-10}$ at the point 
where the intensity at 2.13\,$\mu$m is ${I_{\rm out}}^{-10} = 10^{-10} \cdot I_{\rm max}$, 
this point is found at approximately ${\vartheta_{\rm out}}^{-10} = 5.8$\,arcsec.

\indent The bottom left panel of Fig.\,\ref{f15} shows a plot of the dust temperature 
as a function of angular distance and in units of the inner radius of the dust shell, 
respectively. The inner CDS radius $r_{\rm in}$ can be derived directly from the 
radial profile, as it is the point at which the dust reaches $T_{\rm in} = 1000$\,K. 
It corresponds to an angular radius of $\vartheta_{\rm in} = 10.5$\,mas.

\indent The bottom right panel of Fig.\,\ref{f15} shows the total optical depth as a 
function of wavelength. As expected, local maxima of $\tau_{\lambda }$ are found for the 
silicate absorption features at 9.7\,$\mu$m (SiO stretching vibrations) and 18\,$\mu$m 
(${\rm SiO_2}$ bending vibrations). 

\indent The stellar radius obtained for the best-fitting model is $\vartheta_{\star} = 1.16$\,mas. 
This corresponds to $R_{\star} \simeq 600\,R_{\odot}$ or 2.79\,AU for the adopted distance of 
$D = 2.38 \pm 0.24$\,kpc. Therefore, the inner radius of the CDS is $R_{\rm in} = 9.1\,R_{\star}$ 
= 25.4\,AU. This value is rather high and results from the very high optical depth leading to high 
dust temperatures, so that closer to the central star the dust grains evaporate (see Men'shchikov 
et al.\ \cite{men02b}). The outer radius of the CDS as defined above is at 
$R_{\rm out}^{-10} = 5 \cdot 10^3\,R_{\star}$ = $1.4 \cdot 10^4\,$AU. From the definition 
of $R_{\rm out}^{-10}$, it follows that $R_{\rm out}^{-10}/R_{\rm in} = 550$, and since 
$I_{\rm out}(R_{\rm out}^{-10}) = 10^{-10}\,I_{\rm max}$, 
only a fraction of the order of $10^{-7}$ of the total flux comes from a region with 
 $r \geq R_{\rm out}^{-10}$. 
This explains why models with $r_{\rm out}/r_{\rm in} > 10^3$ do not show much of a change 
with rising $r_{\rm out}/r_{\rm in}$. The value obtained under this assumption is of the same 
order as that found by  Herman \& Habing (\cite{her85}) and therefore seems to give a reliable 
lower limit for the outer diameter of the dust shell. They derived $R_{\rm S} = 4.37 \pm 0.42 
\cdot 10^{16}$\,cm = $2.92 \pm 0.28 \cdot 10^3\,$AU = $6.3 \cdot 10^5\,R_{\odot} \simeq 10^3\,R_{\star}$. 
However, the modeling as well as the observations are both fully allegeable with a dust 
shell that extends from $R_{\rm in}$ to infinity. 

\indent From the observations, the total luminosity can be derived according to 
\begin{equation}
\frac{F_{\rm bol}}{[10^{-10}\,W/{\rm m^2}]} = 3.21 \cdot \frac{L}{[10^4\,L_{\odot}]} 
\cdot \bigg(\frac{D}{[\rm kpc]}\bigg)^{-2} .
\end{equation}

This leads to a luminosity of $L_{\rm obs} = (1.23 \pm 0.12) \cdot 10^4\,L_{\odot}$ 
for the bolometric flux at $\Phi_{\rm ISO}$
which is in very good agreement with the value $L_{\rm mod} = 1.262 \cdot 10^4\,L_{\odot}$
obtained from our best-fitting model using the Stefan-Boltzmann law. 

According to the mass-luminosity relation by Ivezi{\' c} et al.\ (\cite{IE95}), this corresponds 
to a central star mass of $M \simeq 1\,M_{\odot}$, which is in line with predictions from stellar 
evolution theory for stars in this evolutionary stage.

\indent Following Ivezi{\' c} et al.\ (\cite{IE95}), the mass-loss rate $\dot{M}$ can be derived from

\begin{equation}
\dot{M}v_{\rm e} = \tau_{\rm F} \cdot \frac{L}{c} \cdot \Big( 1 - \frac{1}{\Gamma}\Big) \qquad ,
\end{equation}
where
\begin{displaymath}
\Gamma = \frac{F_{\rm rad}}{F_{\rm grav}}
\end{displaymath}
is the ratio of radiation pressure $F_{\rm rad}$ and gravitational force $F_{\rm grav}$ per 
unit volume in the envelope. With $\tau_{\rm F} = 3.2$, $\Gamma = 1.7 \cdot L$ from our 
best-fitting model and the adopted outflow velocity $v_{\rm e} = 15$\,km/s from te Lintel Hekkert 
et al.\ (\cite{TLH91}), $\dot{M} = 2.18 \cdot 10^{-5}\,M_{\odot}$/yr is obtained. 

\subsection{Parameter variations}
 
In this section, we discuss the impact of several input parameters on the process of 
radiative transfer modeling. The best-fitting model (see Sect.\,4.3) is used as a basis 
for the discussion, and the effect of the variations of the parameters in question is outlined. 
The discussion is divided into subsections, in which the change of each parameter is 
discussed separately. To maintain the presentation as concise as possible, the variations 
for minor parameters like $a_{\rm min}$ and the discussion of more complex models (e.g. broken 
power law density distributions, multiple discrete dust shells), that did not deliver better 
fits than the models with less assumptions are omitted. 

\subsubsection{Dust types}

\begin{figure*}
      \begin{center}
\resizebox{80mm}{!}{\includegraphics[angle=-90]{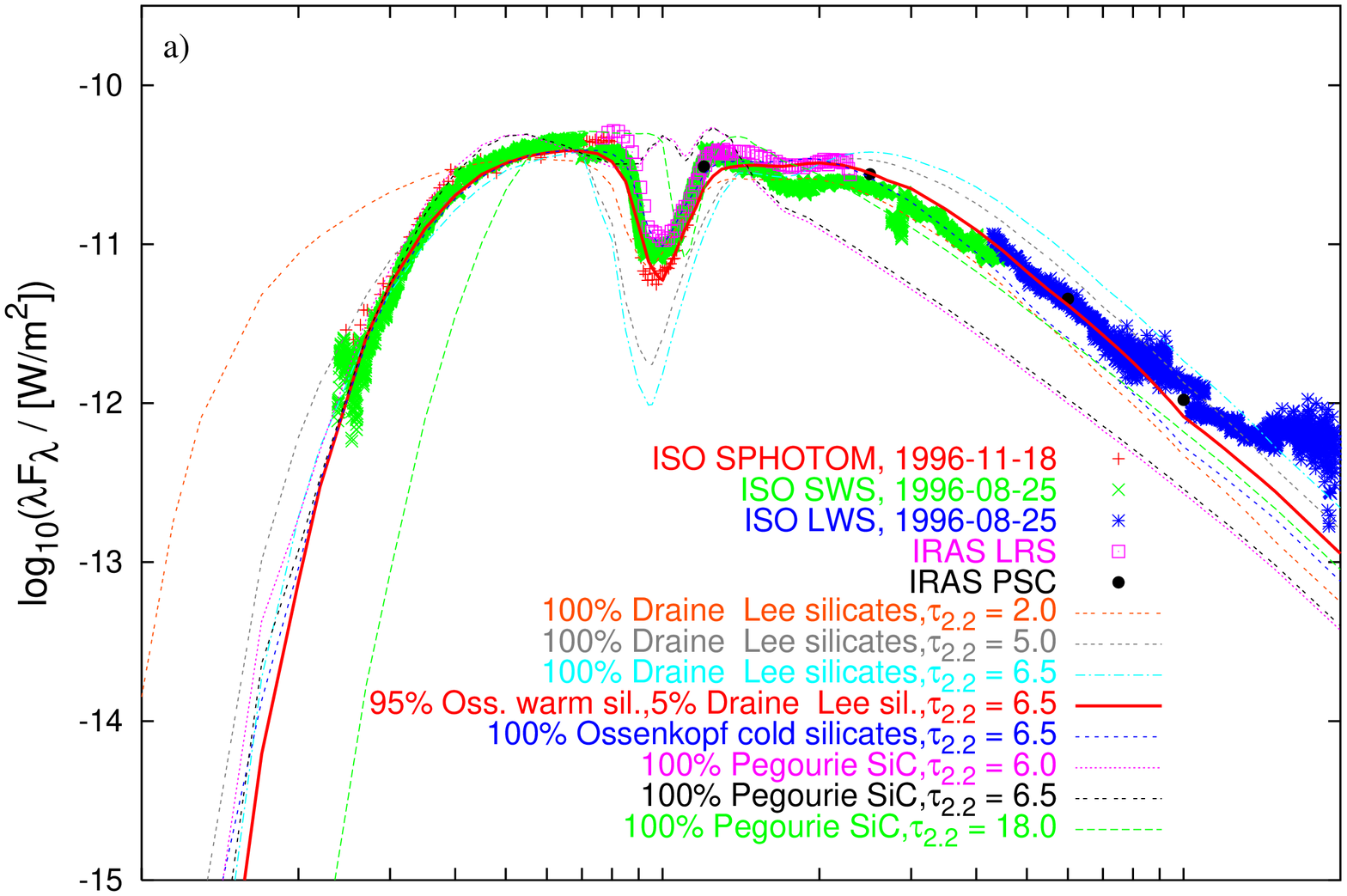}}
\resizebox{80mm}{!}{\includegraphics[angle=-90]{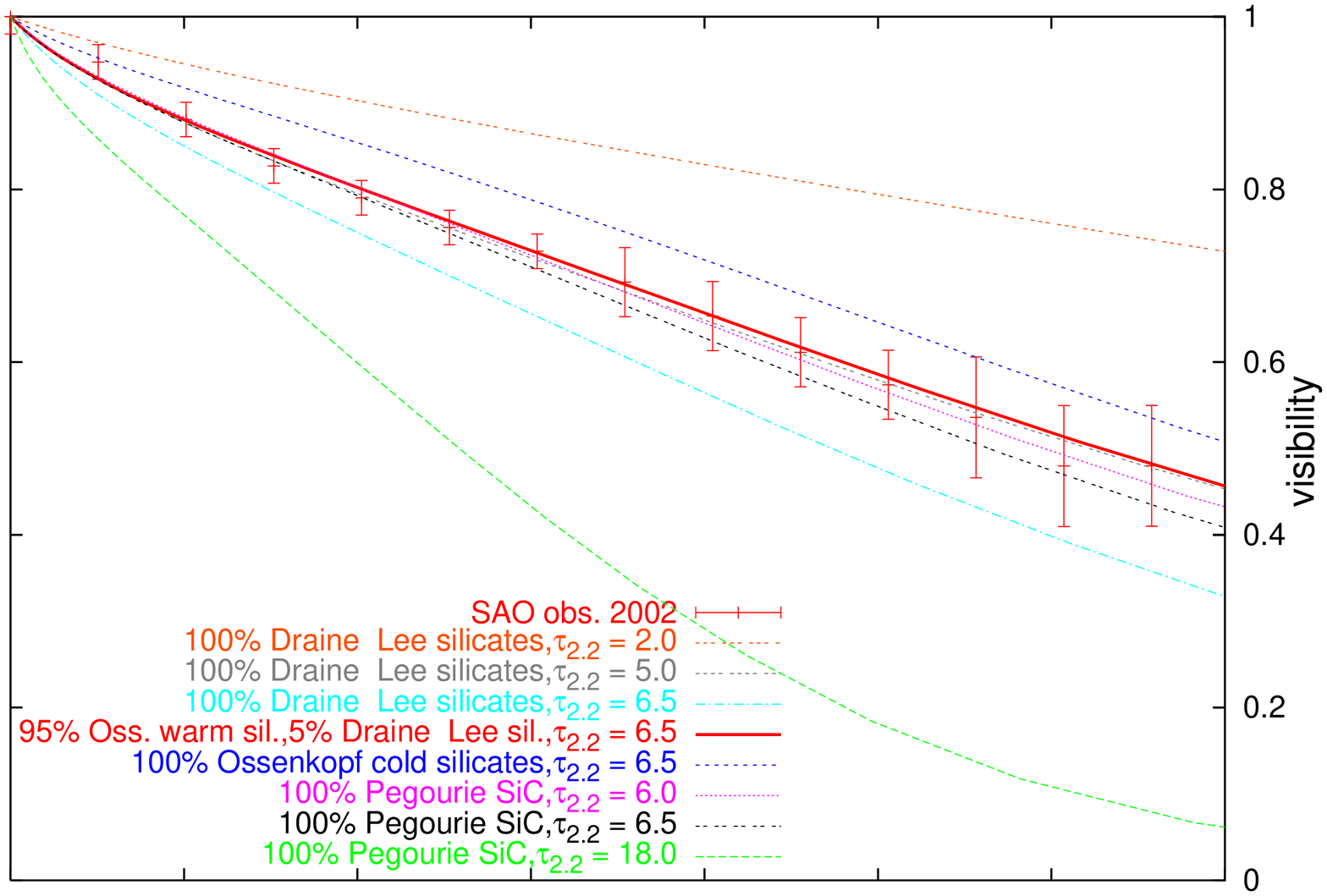}}
\resizebox{80mm}{!}{\includegraphics[angle=-90]{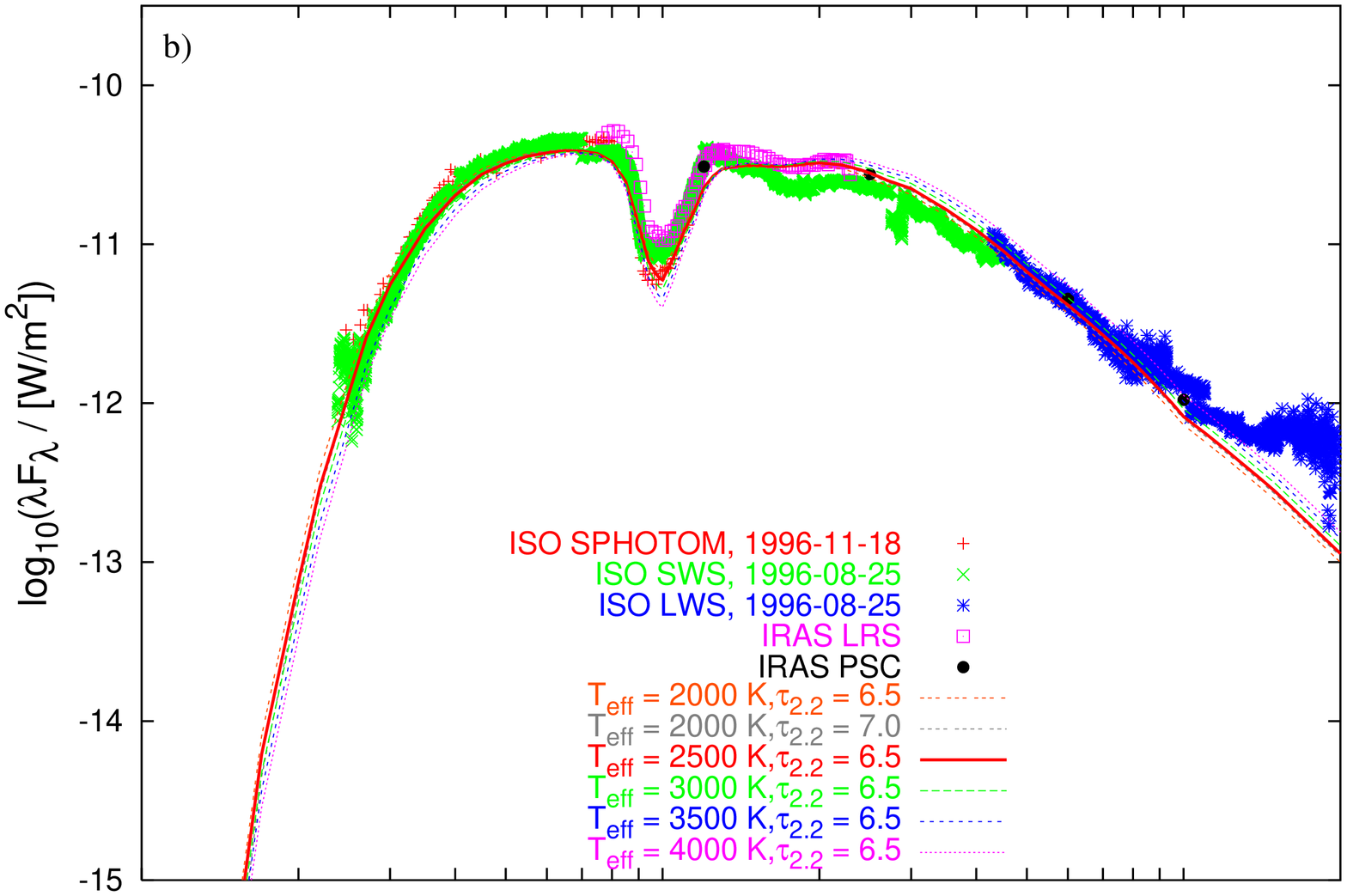}}
\resizebox{80mm}{!}{\includegraphics[angle=-90]{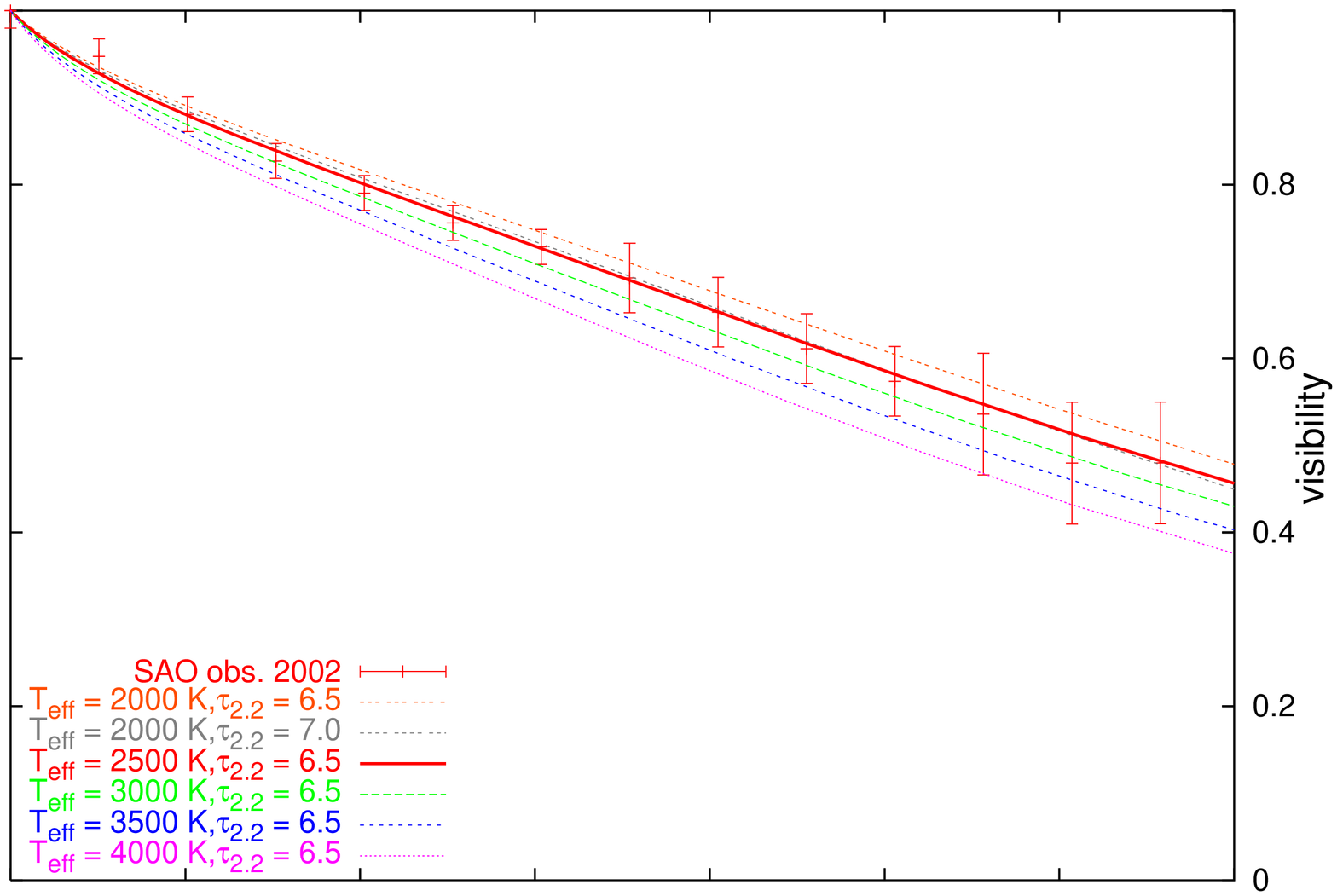}}
\resizebox{80mm}{!}{\includegraphics[angle=-90]{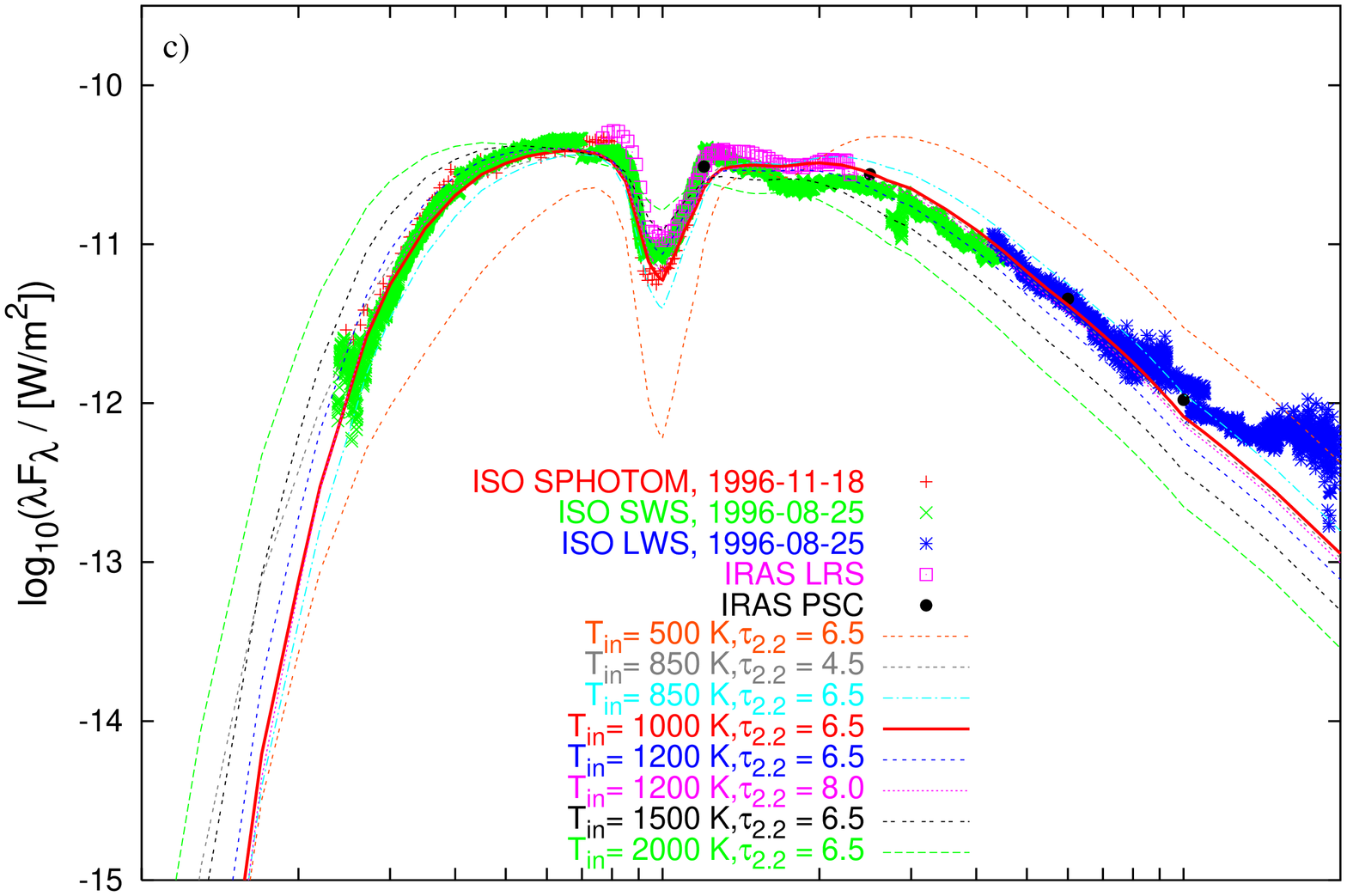}}
\resizebox{80mm}{!}{\includegraphics[angle=-90]{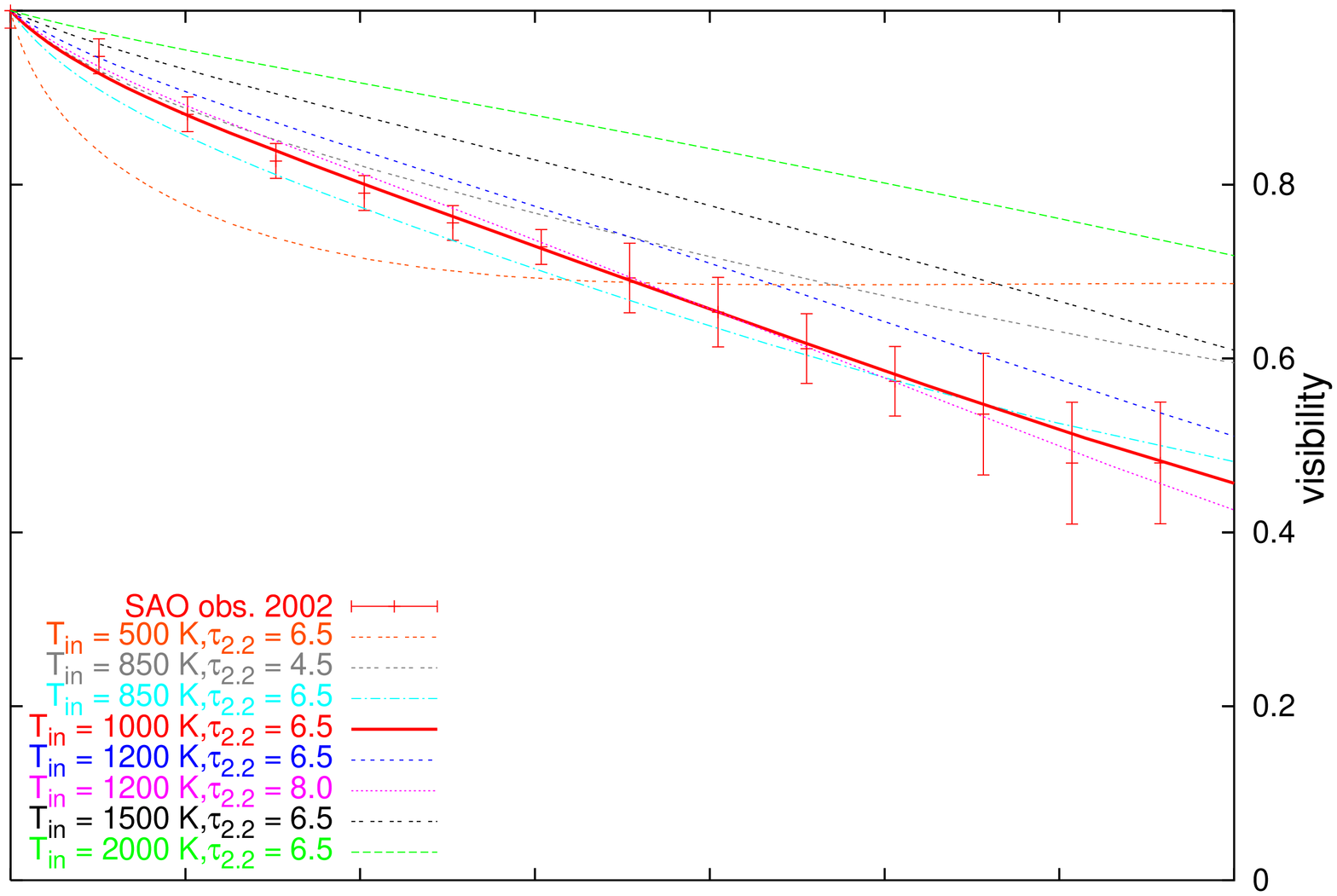}}
\resizebox{80mm}{!}{\includegraphics[angle=-90]{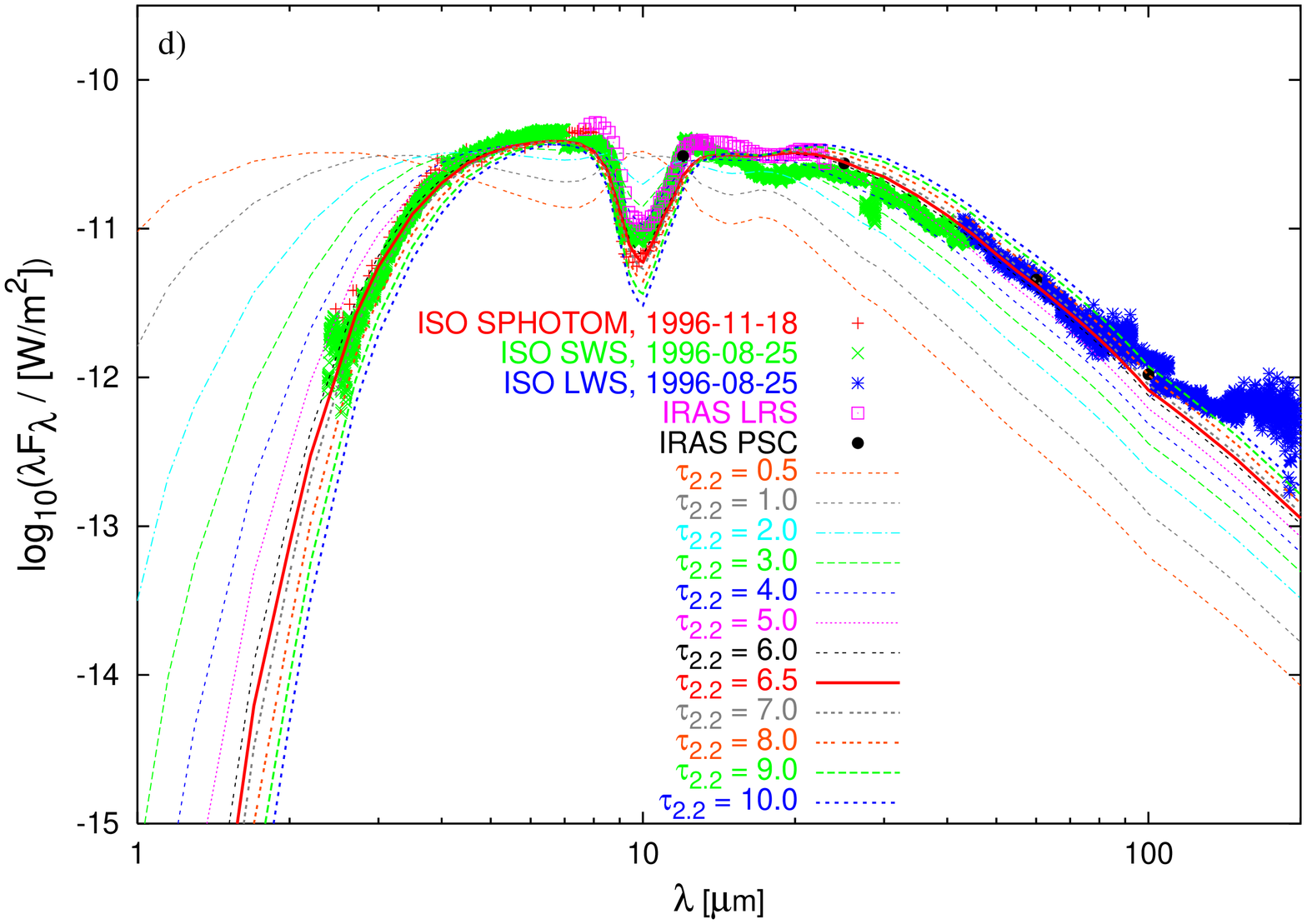}}
\resizebox{80mm}{!}{\includegraphics[angle=-90]{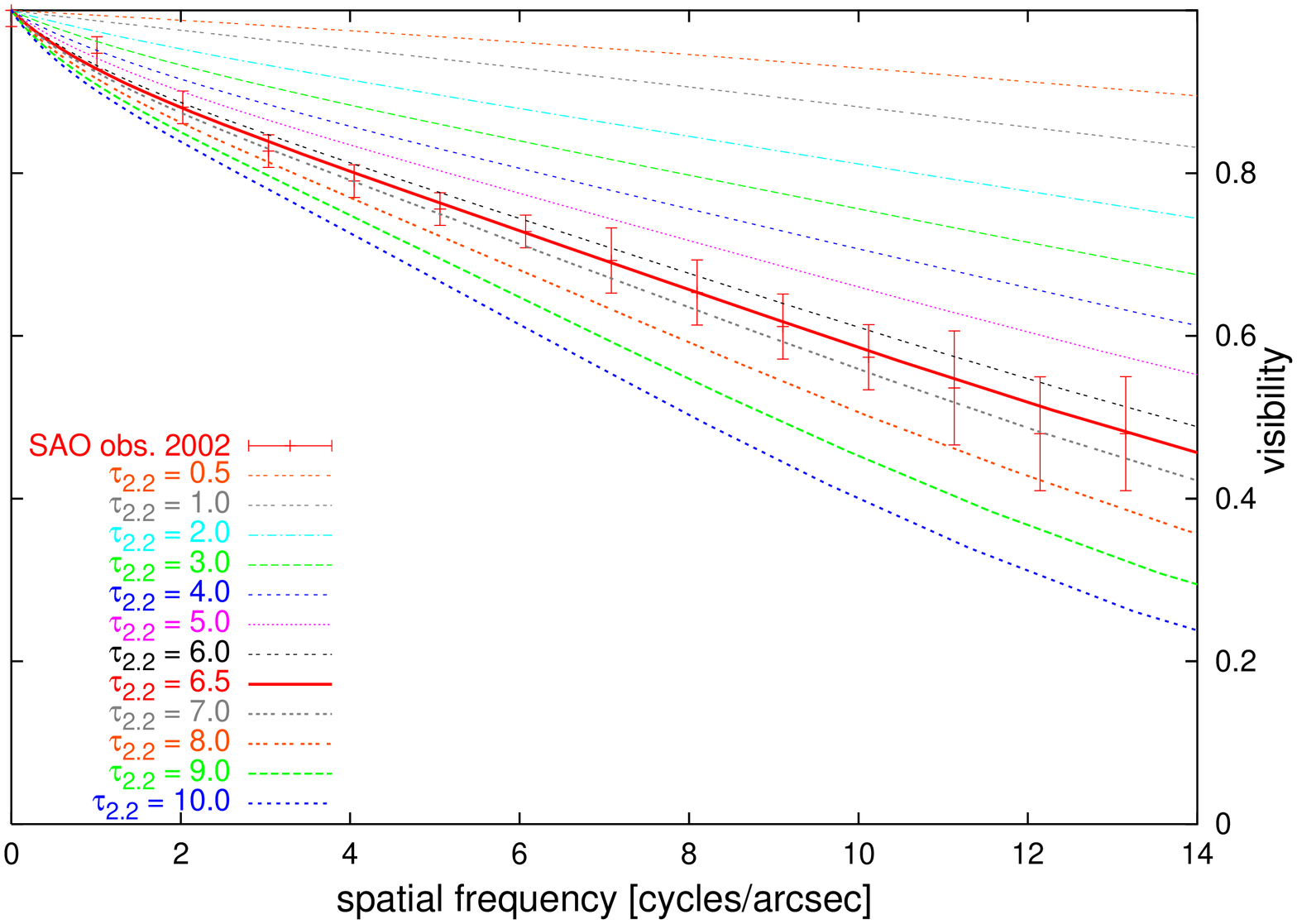}}
      \end{center}
\vspace*{-5mm}

\caption{
SED ({\bf left}) and $K$-band visibility ({\bf right}) of \object{OH\,104.9+2.4} for the 
best-fitting model (thick solid line, 95\,\% OHM warm silicates, 5\,\% DL silicates)
and the impact of several parameter variations.
{\bf a}: Variation of the dust composition. Apart from the
best-fitting model models with other dust types and values of $\tau_{2.2\,\mu{\rm m}}$ 
(see labels) are shown. All other model parameters are fixed to the values 
of the best-fitting model. The models for DL silicates with $\tau_{2.2\,\mu{\rm m}} = 2.0$ and 
$\tau_{2.2\,\mu{\rm m}} = 5.0$ are given as examples for models that fit the SED and visibility 
with this specific dust composition. The same comparison is shown for the Pegourie (\cite{peg88}) 
SiC grains with $\tau_{2.2\,\mu{\rm m}} = 6.0$ and $\tau_{2.2\,\mu{\rm m}} = 18.0$, respectively.
{\bf b}: 
Variation of $T_{\rm eff}$. The best-fitting model is the 
one with $T_{\rm eff} = 2500$\,K. For comparison, a model with $T_{\rm eff} = 2000$\,K 
and $\tau_{2.2\,\mu{\rm m}} = 7.0$ is shown, to illustrate that changing the optical 
depth can make up for the variations caused by $T_{\rm eff}$ (see text for further details).
{\bf c}:
Variation of $T_{\rm in}$. The best-fitting model is labeled 
with $T_{\rm in} = 1000$\,K. For comparison, models with $T_{\rm eff} = 850$\,K and 
$\tau_{2.2\,\mu{\rm m}} = 4.5$ as well as $T_{\rm eff} = 1200$\,K and 
$\tau_{2.2\,\mu{\rm m}} = 8.0$ are shown, to illustrate that a change of the optical 
depth can also make up for the variations caused by $T_{\rm in}$.
{\bf d}:
Variation of $\tau_{2.2\,\mu{\rm m}}$. The best-fitting model 
is labeled with $\tau_{2.2\,\mu{\rm m}} = 6.5$.
}
   \label{f6}
\end{figure*}

One of the most important quantities that have to be fixed in the modeling process 
is the chemical composition of the dust itself. Here, four different dust types are under 
discussion, namely {\it (i)} that of the best-fitting model, which is mainly the warm, 
oxygen-deficient silicates from Ossenkopf, Henning \& Mathis\ (\cite{oss92}), {\it (ii)} 
the type given by the same authors as cold, oxygen-rich  silicates, {\it (iii)} the 
astronomical silicates by Draine \& Lee (\cite{DL84}), which are also included as a small 
fraction in the dust type of our best-fitting model (5\,\%), and finally {\it (iv)} the 
$\alpha$-SiC grains tabulated by Pegourie (\cite{peg88}). The results of radiative transfer 
modeling obtained for these different grain types, with all other parameters fixed to the 
value of the best-fitting model, are presented in Fig.\,\ref{f6}a. The models with OHM cold 
silicates give a slightly worse fit to the data than the models with the warm silicates. 
In general, the DL silicates provide good model fits, though these are not as good as those 
provided by the OHM warm silicates. The small fraction of DL silicates was added to the dust 
composition of the best-fitting model to correctly reproduce the depth of the 9.7\,$\mu$m 
silicate feature. With all other model parameters fixed to the values of the best-fitting 
model (see Sect. 4.3), we need $\tau_{2.2\,\mu{\rm m}} = 2.0$ to fit the SED for DL silicates, 
implying excessive flux at the shorter wavelengths, but $\tau_{2.2\,\mu{\rm m}} = 5.0$ for a 
good visibility fit. The Pegourie SiC composition seems to be unsuitable, as no good 
overall fit to SED and visibility can be found. It is possible to reproduce the depth of the 
silicate feature with an optical depth of $\tau_{2.2\,\mu{\rm m}} = 18.0$, but with a great 
vertical offset (which clearly is due to the fact that SiC has vibrational modes different 
from SiO), and the visibility can be fitted with $\tau_{2.2\,\mu{\rm m}} = 6.0$. 

\subsubsection{Effective temperatures}

Figure\,\ref{f6}b shows the variations of the model SED and visibility with the effective 
temperature of the central star. For a fixed optical depth, a higher value for $T_{\rm eff}$ 
leads to lower flux at both shorter wavelengths and at the 9.7\,$\mu$m silicate feature, and 
higher flux at the long-wavelength tail. The changes are qualitatively similar to those 
introduced by a variation of $T_{\rm in}$ (see next subsection) but much weaker due to the 
rather large $\tau_{2.2\,\mu{\rm m}}$ considered here. The main reason for the changes 
described is the shift of the peak-intensity wavelength of the stellar black-body spectrum 
which induces changes in the radial temperature profile of the dust: the dust temperature 
falls of more steeply for higher effective temperatures. The effect on the visibility is 
rather small in slope and curvature and can be compensated by a small change in 
$\tau_{2.2\,\mu{\rm m}}$, as shown for $T_{\rm eff} = 2000$\,K in Fig.\,\ref{f6}b. 
The visibilities for the model with $\tau_{2.2\,\mu{\rm m}} = 6.5$ and 
$T_{\rm eff} = 2500$\,K and the one with $\tau_{2.2\,\mu{\rm m}} = 7.0$ and 
$T_{\rm eff} = 2000$\,K are almost identical.

\subsubsection{Temperatures at the inner boundary}

Figure\,\ref{f6}c shows the model variations with changing temperature $T_{\rm in}$ 
at the inner dust shell boundary. For $\tau_{2.2\,\mu{\rm m}}$ as high as in our 
favored model, decreasing $T_{\rm in}$ lowers the flux at shorter wavelengths, 
increases the depth of the 9.7\,$\mu$m silicate feature, and raises the flux at the 
long-wavelength tail. This behavior is similar to the changes produced by increasing 
$\tau_{2.2\,\mu{\rm m}}$. Therefore, changes in the SED due to a variation of 
$T_{\rm in}$ can, in principle, be compensated by a corresponding variation of 
$\tau_{2.2\,\mu{\rm m}}$, as illustrated in Fig.\,\ref{f6}c for $T_{\rm in} = 850$\,K 
and $T_{\rm in} = 1200$\,K. On the other hand, the depth of the silicate feature 
constrains the choice of the optical depth. For the visibility function, lowering 
$T_{\rm in}$ leads to a steeper decline at low spatial frequencies and a stronger 
curvature, and therefore a shallower fall-off at high spatial frequencies. With an 
adjusted optical depth (as shown for $T_{\rm in} = 850$\,K and $T_{\rm in} = 1200$\,K), 
the steepness rises, and the curvature decreases with increasing $T_{\rm in}$. Finally, 
it can be concluded that even if models of similar quality to the best-fitting model 
can be found with different $T_{\rm in}$, the value of $T_{\rm in}$ will be close to 1000\,K.

\subsubsection{Reference optical depths}

The optical depth $\tau$ is the main parameter in this study. The variation of this 
quantity has a strong impact on both the SED and the visibility (see Fig.\,\ref{f6}d). 
For the SED, the flux at short wavelengths for low optical depths is relatively high, 
whereas at long wavelengths it is relatively low. In addition, the SiO feature at 
9.7\,$\mu$m is found in emission at the lowest values for $\tau_{2.2\,\mu\rm{m}}$. 
For high optical depths, more flux is reprocessed to long wavelengths as a result of 
the stronger reddening effects. The most important constraint for this parameter is 
the depth of the 9.7\,$\mu$m absorption feature, which is too deep for very high optical 
depths. The visibility mostly keeps its curvature, but steepens for higher optical depths 
due to the stronger obscuration of the central source. The steepening shows a roughly 
linear dependence on the optical depth in the domain under consideration. The best value 
for a simultaneous SED and visibility model is found to be $\tau_{2.2\,\mu\rm{m}} = 6.5$.

\subsubsection{Grain sizes}

\begin{figure*}
      \begin{center}
\resizebox{82mm}{!}{\includegraphics[angle=-90]{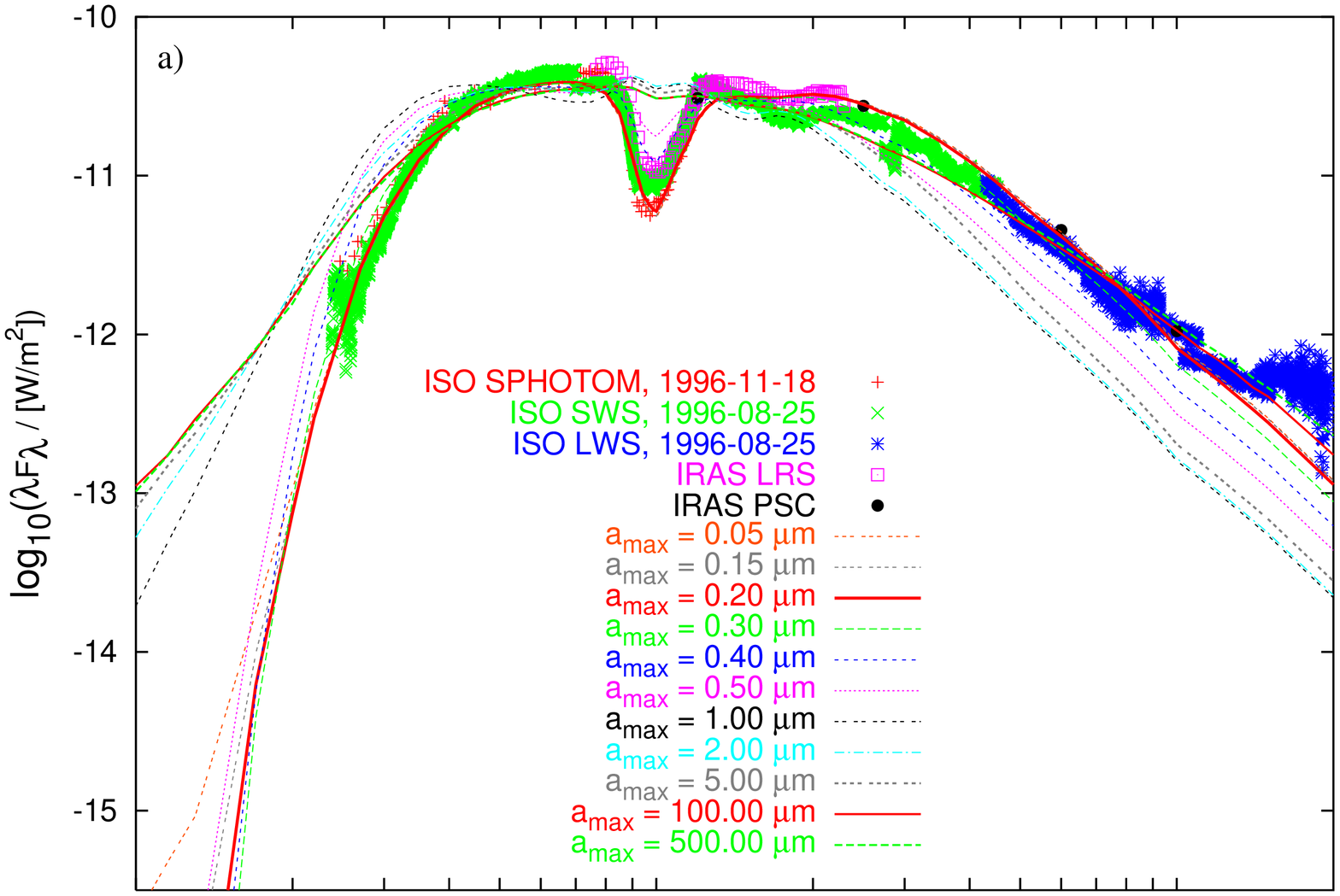}}
\resizebox{82mm}{!}{\includegraphics[angle=-90]{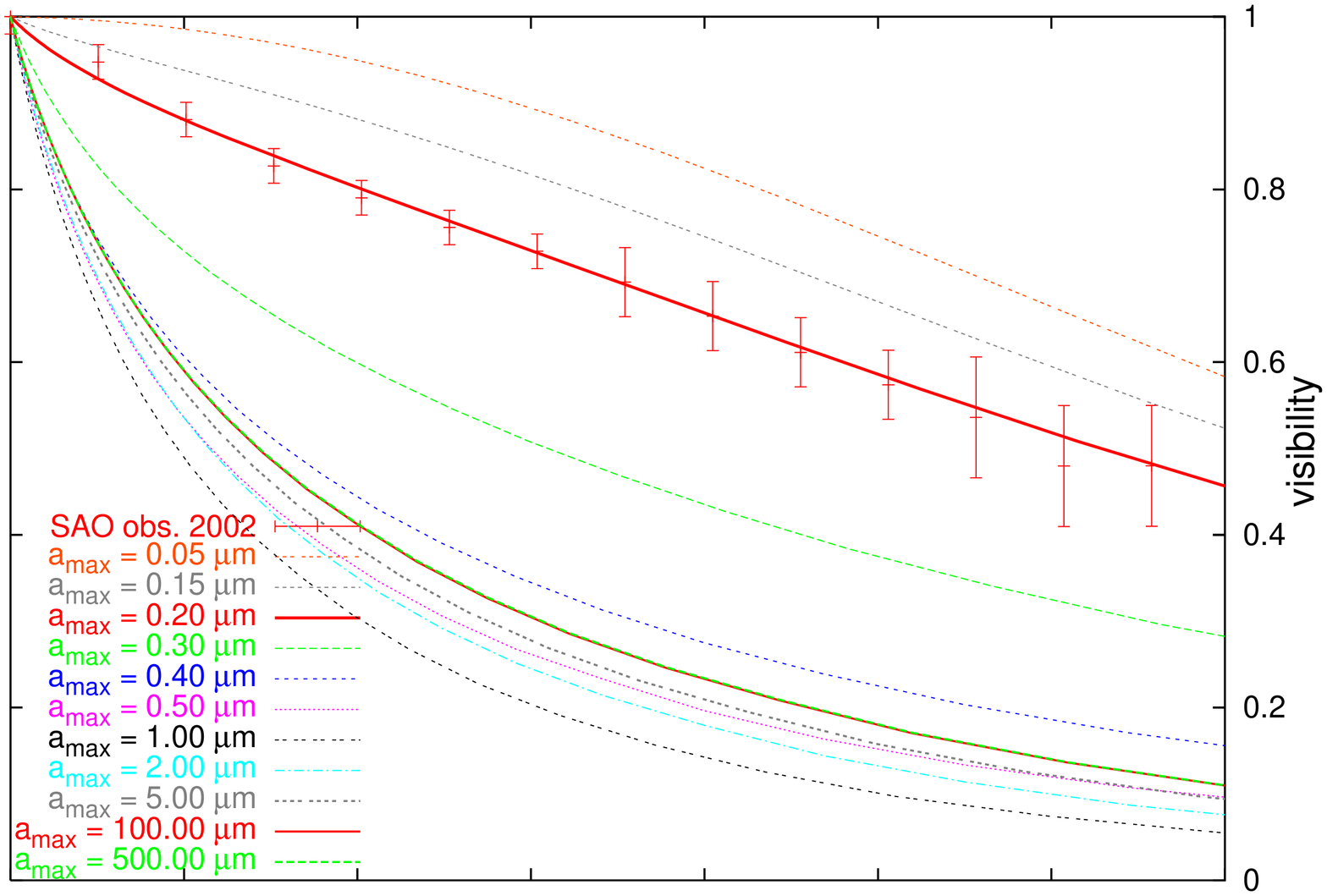}}
\resizebox{82mm}{!}{\includegraphics[angle=-90]{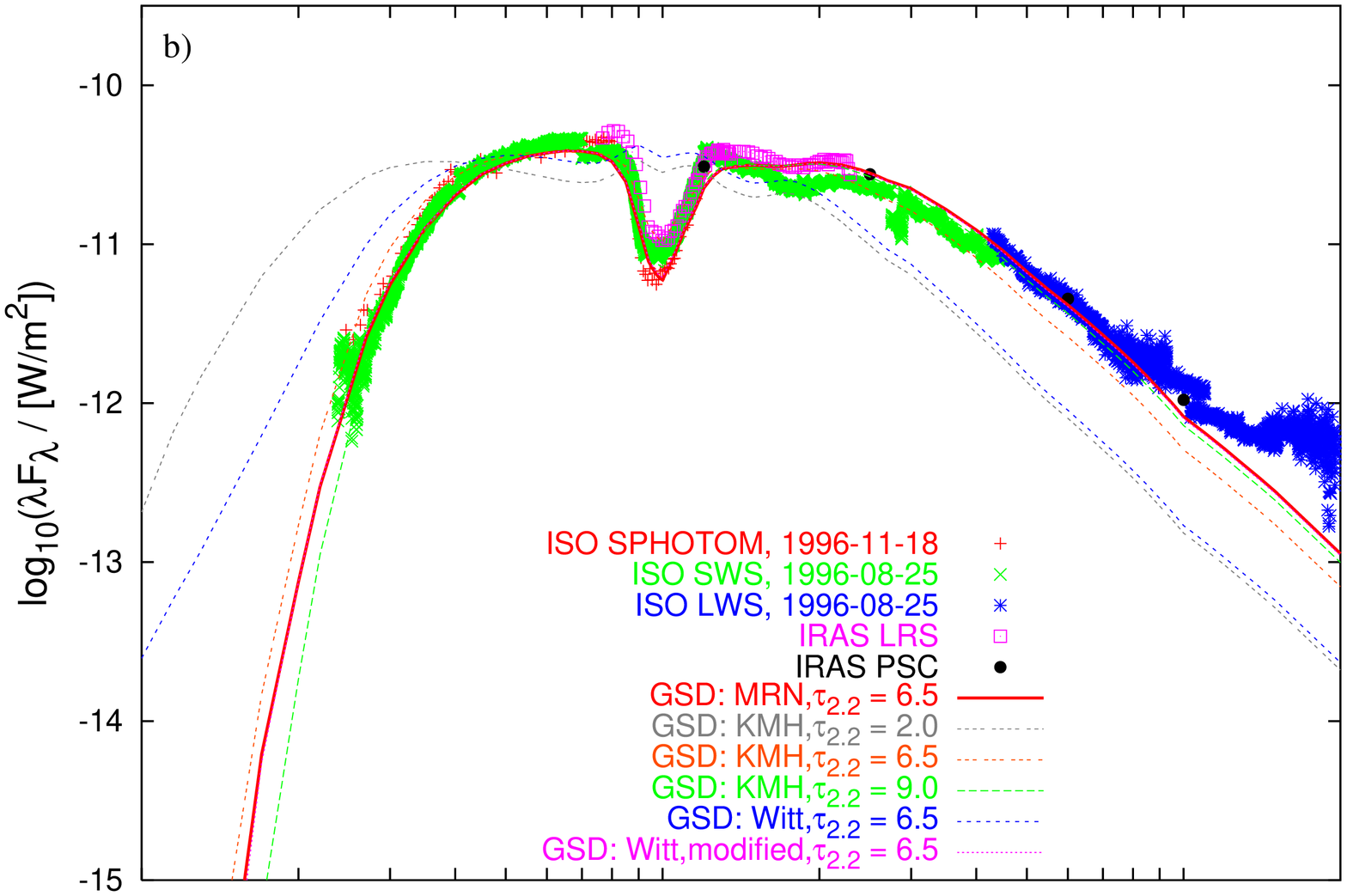}}
\resizebox{82mm}{!}{\includegraphics[angle=-90]{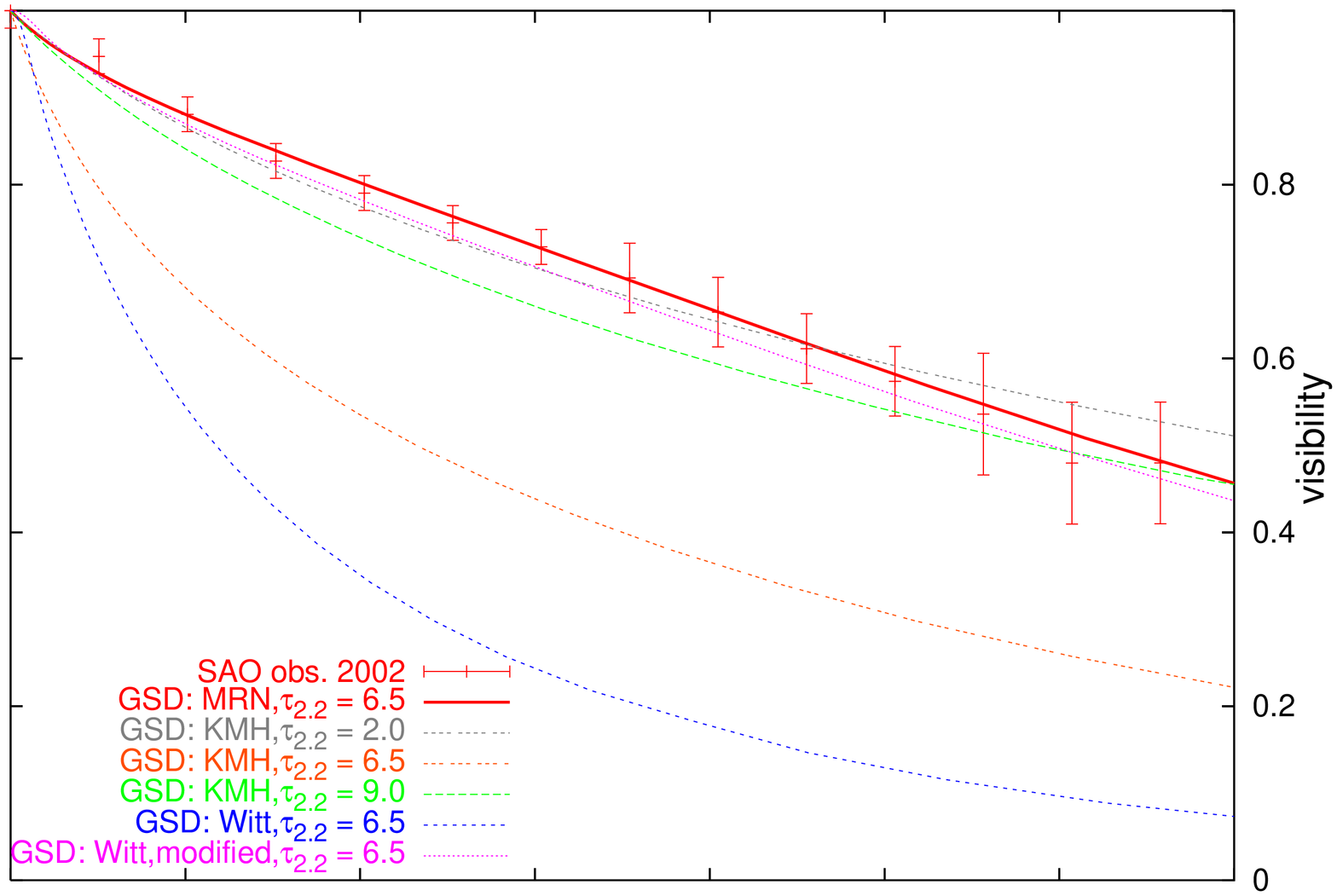}}
\resizebox{82mm}{!}{\includegraphics[angle=-90]{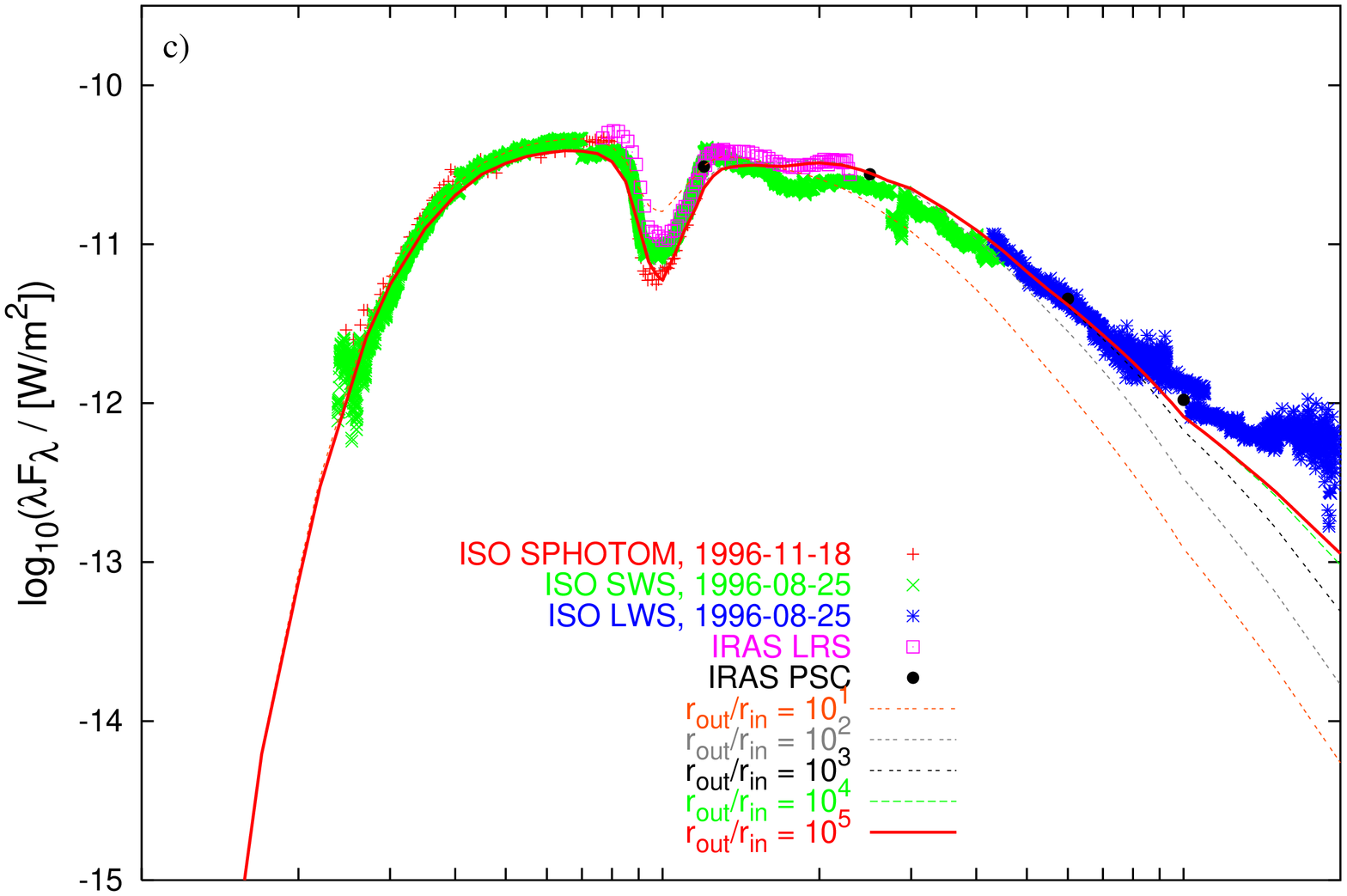}}
\resizebox{82mm}{!}{\includegraphics[angle=-90]{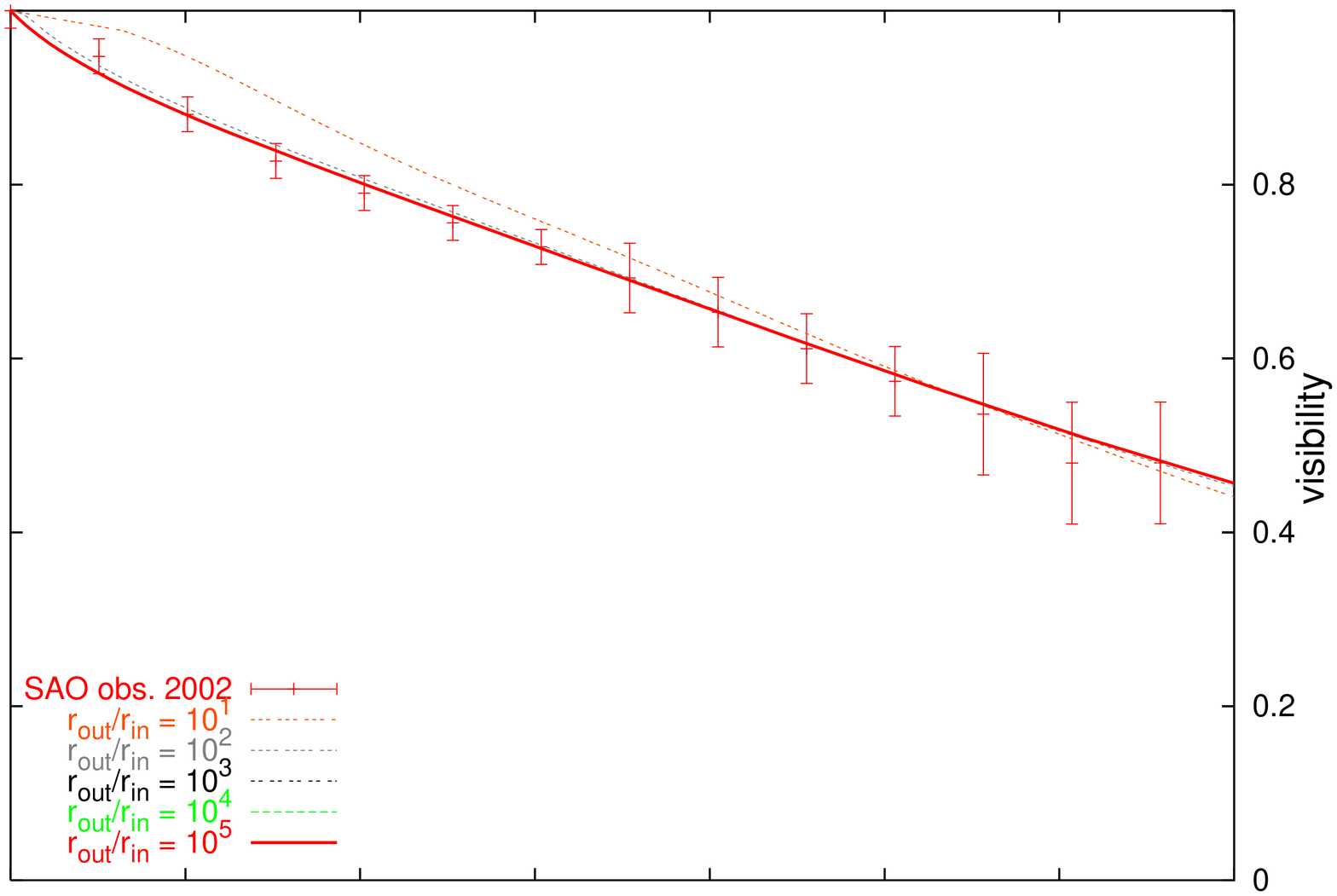}}
\resizebox{82mm}{!}{\includegraphics[angle=-90]{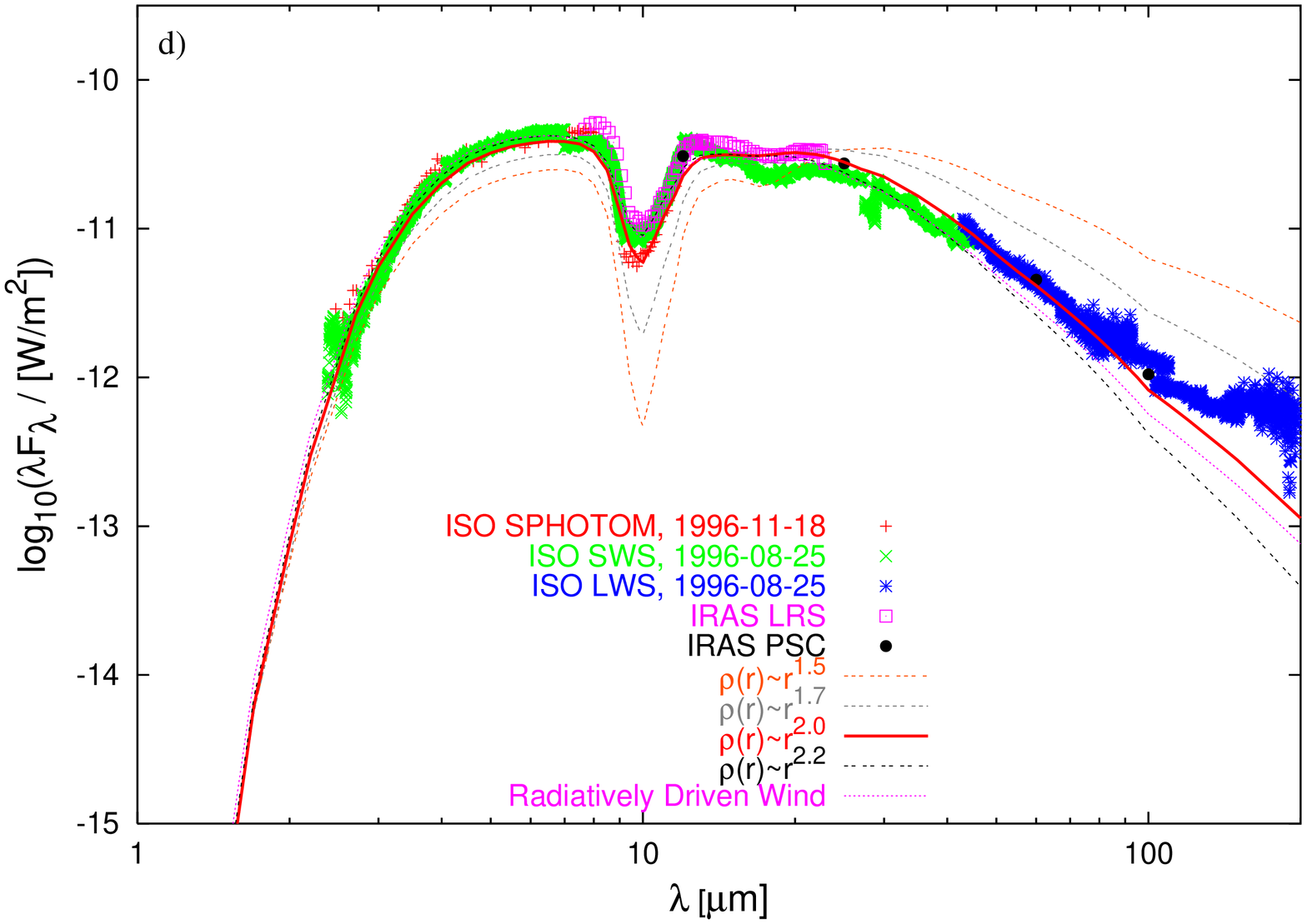}}
\resizebox{82mm}{!}{\includegraphics[angle=-90]{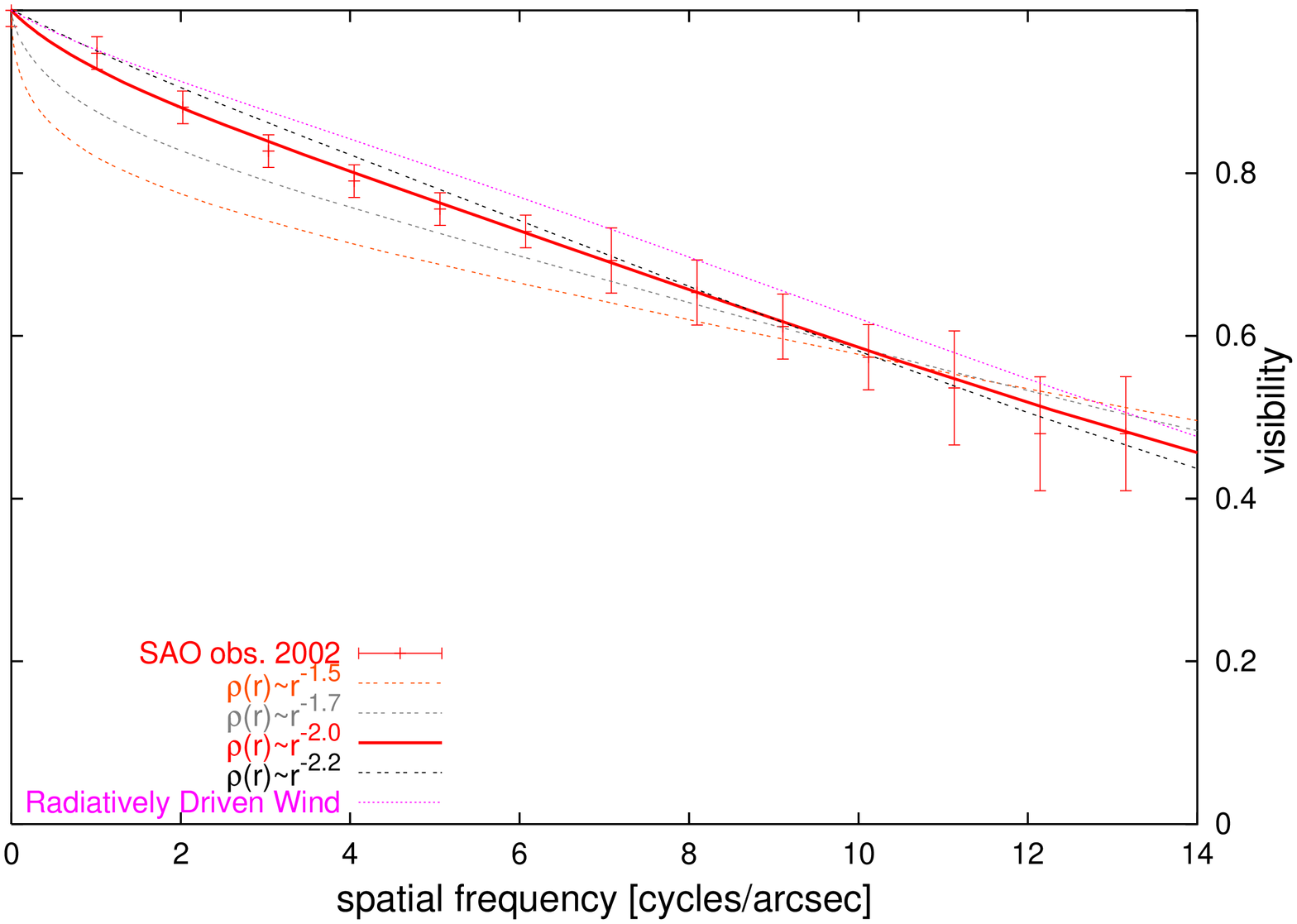}}
      \end{center}
\vspace*{-4mm}

\caption{
SED ({\bf left}) and $K$-band visibility ({\bf right}) of \object{OH\,104.9+2.4} for the 
best-fitting model (thick solid line, 95\,\% OHM warm silicates, 5\,\% DL silicates)
and the impact of several parameter variations.
{\bf a}:
Variation of the maximum grain size $a_{\rm max}$. 
The best-fitting model is the one with $a_{\rm max} = 0.20\,\mu$m.
{\bf b}:
Variation of the grain size distribution. The best-fitting 
model is the one with the MRN distribution. For comparison, models with the KMH 
distribution and $\tau_{2.2\,\mu{\rm m}} = 2.0$ and $\tau_{2.2\,\mu{\rm m}} = 9.0$ 
are plotted as examples of good visibility and SED fits, respectively.
{\bf c}:
Variation of the dust shell thickness. The best-fitting model 
is the one with $\frac{r_{\rm out}}{r_{\rm in}} = 10^5$.
{\bf d}:
Variation of the slope of the radial density distribution. The best-fitting 
model is the one with $\rho(r) \propto r^{-2.0}$. In addition to the
$\rho\propto r^{-n}$ distributions a distribution based on a radiatively
driven wind model is shown (see text for details).
}
   \label{f7}
\end{figure*}

\begin{figure}
      \begin{center}
\resizebox{8.95cm}{!}{\includegraphics[angle=-90]{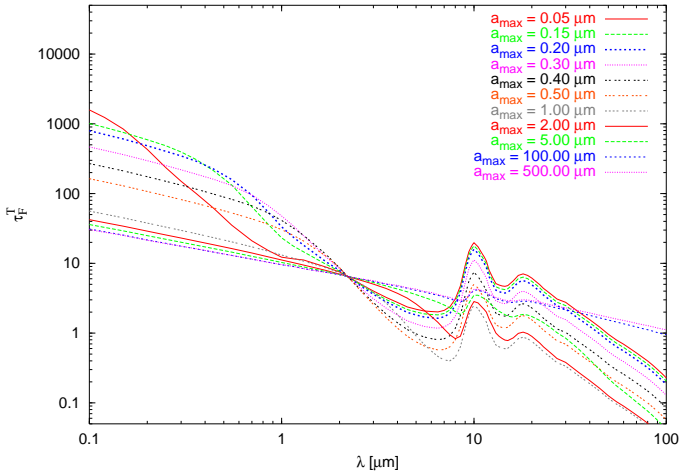}}
      \end{center}
\vspace*{-5mm}

\caption{
Total optical depth spectrum for different maximum grain sizes (see labels).
}
   \label{f8}
\vspace*{-3mm}

\end{figure}

In Fig.\,\ref{f7}a, models assuming an MRN grain-size distribution with different 
maximum grain size $a_{\rm max}$ are shown. All other parameters are fixed to the 
values of the best-fitting model presented in Sect.\,4.3. Concerning the SED, 
$a_{\rm max}$ provides a strong constraint for the scattering and absorption efficiency. 
The averaged (over the size distribution) scattering efficiency per unit volume of the 
grains $Q_{\rm sca}/\hat{V} \propto a^3$ steeply declines with increasing wavelength, 
and can be neglected above a certain wavelength, depending on $a_{\rm max}$ (for an 
example, we refer to Fig.\,\ref{f15}, top left panel of the best-fitting model). 
The size-averaged absorption efficiency depends on the grain size only at short 
wavelengths and becomes independent of $a_{\rm max}$ when $\lambda \gg a$ (see 
Fig.\,\ref{f8}). Therefore, the fluxes observed at the shortest wavelengths place 
an upper limit on the maximum grain size. The depth of the 9.7\,$\mu$m silicate 
absorption feature itself leads to the prediction that $a_{\rm max} \leq 0.50\,\mu$m 
for the current configuration. In the long wavelength regime ($\lambda \geq 100\,\mu$m), 
it is possible to find a model that fits better than the best-fitting model, but the 
improvement of the fit quality is relatively small. The slightly better fit might be 
due to a fixed $F_{\rm bol}$ in combination with the fact that the models in question 
become optically thin at short wavelengths (Fig.\,\ref{f8}), since the large grains 
engross large fractions of the dust mass. In this case, a large fraction of $F_{\rm bol}$ 
is emitted at short wavelengths, resulting in lower flux at long wavelengths. In addition, 
the problems mentioned in Sect. 4.3.1 may play a role. 

\indent The $K'$-band visibility is very sensitive against scattering and thus depends 
strongly on the assumed grain sizes. Increasing $a_{\rm max}$ has a similar effect to 
increasing the optical depth and results in a stronger decline of the visibility function. 
As an MRN-like distribution declines very steeply towards larger grain sizes, most of the 
particles in the model remain small when $a_{\rm max}$ is increased. But, due to the strong 
dependence of the scattering and absorption properties on the grain sizes at 2.12\,$\mu$m, 
the impact of these few extra large grains on the visibility is comparable to that of the 
large bulk of small grains (Kr\"uger \& Sedlmayr \cite{ks97}). The curvature of the 
visibility changes its sign somewhere around $a_{\rm max} \leq 0.20\,\mu$m. This behavior 
reflects the change of the spatial intensity distribution. At large radial offsets $b$ 
from the central star, $I(b) \propto b^{-3}$ holds, since the optical depth along the line 
of sight becomes small (Jura \& Jacoby \cite{jj76}). 

\indent Regarding the wavelength dependence of the total optical depth, the choice of 
$a_{\rm max}$ is only of minor importance beyond a certain range of values (see Fig.\,\ref{f8}). 
For $a_{\rm max} \leq 1\,\mu$m and wavelengths larger than 5\,$\mu$m, the impact of the 
grain size is limited. The shape of the function is almost the same, and the offset can 
be fixed by adjusting the optical depth of the models. If the grains become too large 
and too few, the function approximately converges towards a power law, since the 
characteristics of the grains then play a minor role. 

\subsubsection{Grain-size distributions}

Instead of the standard MRN grain-size distribution, several other GSD functions 
can be adopted for the dust under examination. In this study, variations of the 
standard MRN function have been tested.

\indent Apart from the MRN distribution, the distribution from Kim, Martin, \& Hendry\ 
(\cite{kmh94}, hereafter KMH) is frequently used. It assumes $n(a) \propto a^{-q} 
\cdot e^{-a/a_0}$, with $q = 3.5$, for $a \geq a_{\rm min}$, where $a_0$ has to be 
fixed for the exponential falloff, which replaces the sharp upper cutoff of the MRN 
distribution. The value chosen here is $a_0 = 0.17\,\mu$m, which should lead to a 
distribution comparable to the standard MRN. The results are shown in Fig.\,\ref{f7}b. 
For $\tau_{2.2\,\mu{\rm m}} = 6.5$, the fit is not good. For a good SED fit, 
$\tau_{2.2\,\mu{\rm m}} = 9.0$ is needed. However, we need $\tau_{2.2\,\mu{\rm m}} = 2.0$ 
for a good visibility fit. Therefore, models with a KMH grain-size distribution provide 
a good fit of either the SED or the visibility, but not both simultaneously.

\indent In addition to KMH we implemented in the DUSTY code the GSD from Witt 
et al.\ (\cite{wit01}). Here, the MRN function remains unchanged, but a certain 
break point $a_{\rm break}$ is introduced, where the slope of the exponential decline 
$q$ changes. We first used the original values $q_1 = 3.5$, $q_2 = 4.0$, $a_{\rm min} 
= 0.005\,\mu$m, $a_{\rm break} = 0.5\,\mu$m and $a_{\rm max} = 2.0\,\mu$m given by 
Witt et al.\ (\cite{wit01}). Obviously, results obtained with this parameter set are 
not directly comparable to the best-fitting model, due to different grain sizes. With 
$a_{\rm break} = 0.15\,\mu$m and $a_{\rm max} = 0.22\,\mu$m, an SED fit of comparable 
quality to our best-fitting model is obtained, while the visibility fit is worse at 
intermediate spatial frequencies. Since reasonable improvements all converge towards 
the parameters of our best-fitting model, 
our assumption about the MRN GSD is justified.

\subsubsection{Dust-shell thickness}

In Fig.\,\ref{f7}c different relative dust-shell thicknesses are examined. Except for 
the model with $\frac{r_{\rm out}}{r_{\rm in}} = 10$, the SED fits are almost the same 
up to $\lambda \simeq 30\,\mu$m, while the model visibility remains essentially unchanged. 
Taking into account the poor visibility fit, dust shells with 
$\frac{r_{\rm out}}{r_{\rm in}} \ll 10^2$ can be ruled out. Above 30\,$\mu$m, the SED 
falls off more shallowly for rising geometrical thickness, providing a better fit to 
the observational data. The additional cold dust introduced by the larger outer boundary 
increases the far infrared flux but does not affect the SED at shorter wavelengths. 
The only difference between the models $\frac{r_{\rm out}}{r_{\rm in}} = 10^4$ and 
$\frac{r_{\rm out}}{r_{\rm in}} = 10^5$ is a small deviation for $\lambda \gtrsim 120\,\mu$m. 
Thus, from our analysis the model with $\frac{r_{\rm out}}{r_{\rm in}} = 10^5$ is slightly 
favored, whereas the relative dust-shell thickness turned out to be a non-crucial 
parameter in the modeling process (see also discussion above about $R_{\rm out}^{-10}$). 

\subsubsection{Density distributions}

So far, it was assumed that $\rho \propto r^{-2}$ holds within the dust shell. 
Assuming the outflow velocity $v_{\rm e}$ is the same as for the standard model 
and constant within the CDS and a constant dust-to-gas ratio, any density 
slope  shallower/steeper than $\rho \propto r^{-2}$ means that the stellar 
mass-loss rate $\dot{M}$ is non-stationary and decreases/increases with time. 
Figure\,\ref{f7}d shows the result of variations in the density distribution. 
Since the overall agreement between observations and models worsens for $\rho 
\propto r^{-x}$ with $x \neq 2$, our standard density profile is justified. 

\indent For AGB stars, the envelope expansion is driven by radiation pressure 
on the dust grains. Therefore, a simple dust density profile $\rho(r) \propto r^{-x}$ 
may not be the most realistic one. If the variation of the flux-averaged opacity 
with radial distance as well as the grain drift is negligible, an analytic solution 
for the problem of a radiatively driven wind can be found (Ivezi{\' c} et al.\ \cite{IE99}) 
which might be more appropiate for modeling cool evolved stars such as 
\object{OH\,104.9+2.4}. In Fig.\,\ref{f7}d such a model is shown, whose other 
parameters are identical to those of our best-fitting model. The SED fit agrees 
with the observations at most of the wavelengths discussed. Only for short wavelengths 
($\lambda < 3\,\mu$m) and long wavelengths ($\lambda > 80\,\mu$m) is there a 
noticeable deviation from the observational data. Unfortunately, no better model 
fit of the long wavelength regime could be obtained compared to our best-fitting 
model without a radiatively driven wind. In terms of the visibility, some models provide 
fits of comparable quality to our best-fitting model shown in Fig.\,\ref{f7}d 
or even slightly better ones (with a worse SED fit). After all, both modeling 
approaches provide models of similar quality with otherwise identical parameters, 
while our best-fitting model is still slightly favoured. 

   \section{Summary and conclusions}\label{conclusions}
We present diffraction-limited 2.12\,$\mu$m speckle interferometric observations of the 
circumstellar dust shell around the highly obscured type II-A OH/IR star \object{OH\,104.9+2.4}. 
The resolution achieved with the SAO 6\,m telescope is 74\,mas, which is sufficient to fully 
resolve the CDS at this wavelength. From an azimuthally averaged 1-dimensional Gaussian fit 
of the visibility function, the diameter was determined to be 47 $\pm$ 3\,mas (FWHM), which 
corresponds to a diameter of 112 $\pm$ 13\,AU for an adopted distance of $D = 2.38 \pm 0.24$\,kpc. 
The reconstructed 2-dimensional visibility shows no major deviation from spherical symmetry. 

\indent In order to investigate the structure and the properties of the CDS, we performed 
1-dimensional radiative transfer calculations to model {\it simultaneously} the spectral 
energy distribution (ISO SWS/LWS spectrum), as well as the 2.12\,$\mu$m visibility function 
from our observations. The obtained dust shell properties correspond to a standard single-shell 
model (uniform outflow). Multiple component models do not match the observations better 
as the simple uniform outflow models, while they are certainly not excluded with the 
remaining uncertainities. Among the $\sim 10^6$ models calculated, the ISO SED, which appears 
to correspond to the minimum pulsation phase of \object{OH\,104.9+2.4}, was best reproduced by 
a model where the inner rim of the dust shell (i.e.\ the dust condensation zone) has a dust 
temperature of $T_{\rm in} = 1000$\,K and an angular diameter of 10.5\,mas corresponding to 
9.1\,$R_{\star}$ (Fig.\,\ref{f15}, bottom left panel). At 2.12\,$\mu$m, this zone is not 
limb-brightened, since the model is optically thick ($\tau_{2.2\,\mu{\rm m}} = 6.5$). 

\indent During the entire model grid calculations, the validity of the parameters given above 
as well as of other parameters such as dust temperatures, grain sizes, and effective temperatures 
was comprehensively investigated. The grain-size distribution was found to agree with 
a standard MRN distribution function, $n(a) \propto a^{-3.5}$, with  $a$ ranging between 
$a_{\rm min} = 0.005\,\mu$m and $a_{\rm max} = 0.2\,\mu$m, which shows that the influence 
of more complex processes like the dust formation process itself on the model seem to be rather 
little. A two-component dust composition consisting of 95\,\% OHM warm silicates and 5\,\% DL 
silicates was found to match the observations best. Assuming a single black-body spectrum for 
the spectral shape of the central source, it is best represented by a model with $T_{\rm eff} = 2500$\,K. 
However, moderate variations of $T_{\rm eff}$  only slightly affect the model results. From the 
best fitting model, we derived a bolometric flux at Earth of $F_{\rm bol} = 6.99 \cdot 
10^{-11}$\,W/$\rm m^2$, in accordance with the ISO observational data for the SED at 
$\Phi_{\rm ISO} = 0.5 = \Phi_{\rm min}$. This bolometric flux corresponds to a luminosity of 
$L = (1.23 \pm 0.12) \cdot 10^4\,L_{\rm \odot}$ at a distance of $D = 2.38 \pm 0.24$\,kpc. 
The radius of the central star is found to be $\vartheta_{\star} = 1.16\,$mas or $R \simeq 
600\,R_{\rm \odot}$, and the central-star mass was derived to be $M \simeq 1\,M_{\rm \odot}$ 
using the method of Ivezi{\' c} et al.\ (\cite{IE95}). Though the density distribution was varied 
greatly, the general $\rho(r) \propto 1/r^2$ dependence was proven to best fit the observations. 
The present study matches the observations for a broad wavelength range (from 1\,$\mu$m to 220\,$\mu$m).

\indent  As regards our visibility at 2.12\,$\mu$m, the difference in pulsation phase between 
the SAO observations and the ISO measurements had to be taken into account. According to the 
observations, the bolometric flux between the 1996 ISO SED phase and the 2002 SAO visibility 
phase differs by a factor of 3.3. Therefore, this factor had to be applied to the bolometric 
flux of the visibility model in order to simultaneously fit the SED and the visibility. 

\indent As concerns the study of the parameter space, the grain size turned out to be the key 
parameter in adjusting the visibility fit while preserving the quality of the SED fit. 
Both the slope and the curvature of the visibility depend sensitively on the assumed grain radii. 
The dust mass-loss rate $\dot{M}$ is well constrained by the shape of the SED at longer 
wavelengths and especially by the shape of the silicate absorption feature. For given optical 
constants, the value of $\dot{M}$, as derived from the match of the feature, is not very 
sensitive to changes in the input parameters. The derived value of $\dot{M} = 2.18 \cdot 
10^{-5}\,M_{\odot}/$yr is of the same order of magnitude as the value of $\dot{M} = 5.58 
\cdot 10^{-5}\,M_{\odot}/$yr given by Heske et al.\ (\cite{hfo90}). The shape of the observed 
visibility and the strength of the silicate feature constrain the possible grain 
radii and optical depths of the model, which are found to be the same values that provide 
the best fit to the SED. 

\indent While our model of \object{OH\,104.9+2.4} simultaneously fits the observed SED and 
visibility very well, the modeling included some simplifications. The central source is not 
a perfect black body, as the dynamical stellar atmosphere exhibits a complex structure 
and heterogeneous chemistry. In addition, the feature at 18\,$\mu$m could not be reproduced 
very well in the SED, due to the simplified dust composition, and the wavelengths above 100\,$\mu$m 
could only be fitted to a certain extent. This is probably due to the fact that DUSTY allows 
for the entry of a grain-size distribution, but the code averages dust properties over the 
entire size distribution before performing any further calculations (see e.g.\ Wolf \cite{wol03}). 
In our case, this leads in particular to an absence of the very large grains found in 
the tail of the grain-size distribution, which have very high and almost constant absorption 
and scattering efficiencies. By choosing larger grain sizes, we obtain a much better SED fit 
for $\lambda \geq 50\,\mu$m, as the emission from the large grains, which have a flatter 
absorption efficiency and a higher scattering efficiency as smaller grains, then traces the 
shape of the SED at the long wavelengths, where the dust shell is otherwise optically thin. 
Unfortunately, this also leads to an optically thin solution for shorter wavelengths due to 
a lack of small grains, as most of the dust mass is in this case utilized by large grains.

\indent For future considerations, it would be very useful to have additional $K$-band photometry 
to secure the derived value for the pulsation phase, as well as photometry in other bands to get 
better insight into the wavelength dependence of the pulsation amplitude. More visibility measurements, 
e.g.\ in the $J$, $H$, $N$, and $V$ bands, would provide additional information on the structure 
of the dust shell, while additional visibilities in the $K'$ band at another pulsation phase 
would allow for more detailed conclusions about the response of the dust shell to the variation 
in the central object, as there may be light-travel time effects involved due to the large scale 
of the CDS itself. 

\begin{acknowledgements}
The observations were carried out with the SAO 6\,m telescope, operated by the Special 
Astrophysical Observatory, Russia. Additional NIR photometric data was provided by B.\ Yudin 
at the CAO, operated by the Sternberg Astronomical Institute of the Moscow State University, 
Russia. We thank the Infrared Space Observatory (ISO) operators at the European Space Agency 
(ESA) for providing the SED data. This research has made use of the SIMBAD database, operated 
by CDS in Strasbourg, as well as the Gezari catalogue, published by NASA, and the NASA Astrophysics 
Data System (ADS), operated by NASA. A.~B.\ M.\ acknowledges support from the Natural Sciences 
and Engineering Research Council of Canada (NSERC). Finally, we thank the referee, Dr.\ Peter 
Woitke, for many helpful comments that stimulated further improvements of the manuscript. 
\end{acknowledgements}


\end{document}